\documentclass[twocolumn,superscriptaddress,altaffilletter,prd]{revtex4-1}

\usepackage[utf8]{inputenc}
\usepackage{lineno,hyperref}
\usepackage{import}
\usepackage{siunitx}
\usepackage{amssymb}

\usepackage{./hepparticles} 
\usepackage[italic]{heppennames}

\modulolinenumbers[5]

\usepackage[dvipsnames]{xcolor} 

\RequirePackage{xspace}

\RequirePackage{hyperref}

\DeclareRobustCommand{\Pgb}{\HepParticle{\beta}{}{}\xspace} 

\def\mus  {\ensuremath{\,\mus}\xspace}

\def\kg   {\ensuremath{\,\mathrfm{kg}}\xspace}
\def\ton  {\ensuremath{{\,\mathrm{t}}}\xspace}

\def\mus        {\ensuremath{\,\mu\mathrm{s}}\xspace}    

\def\to                 {\ensuremath{\rightarrow}\xspace}

\def\gsim{{~\raise.15em\hbox{$>$}\kern-.85em
          \lower.35em\hbox{$\sim$}~}\xspace}
\def\lsim{{~\raise.15em\hbox{$<$}\kern-.85em
          \lower.35em\hbox{$\sim$}~}\xspace}

\newcommand{\QBB}{\ensuremath{Q_{\!\Pgb\kern-0.1em\Pgb}}\xspace}

\newcommand{\IKreF}{\HepIsotope{Kr}{}{85}\xspace}

\newcommand{\IBe}{\HepIsotope{B}{}{8}\xspace}

\newcommand{\Dnuenue}{\hbox{\Pgn\kern-0.3em\Pe\kern-0.3em$\to$\kern-0.1em\Pgn\kern-0.3em\Pe}\xspace}
\newcommand{\ISmofS}{\HepIsotope{Sm}{}{147}\xspace}

\newcommand{\IGd}{\HepIsotope{Gd}{}{}\xspace}

\newcommand{\IGdoFt}{\HepIsotope{Gd}{}{152}\xspace}

\newcommand{\ICof}{\HepIsotope{C}{}{14}\xspace}

\newcommand{\Dznbb}{\hbox{0\Pgn\kern-0.3em\Pgb\kern-0.35em\Pgb}\xspace}
\newcommand{\Dtnbb}{\hbox{2\Pgn\kern-0.3em\Pgb\kern-0.35em\Pgb}\xspace}

\newcommand{\COt}{\HepChemical{O}{2}{}\xspace}
\newcommand{\COtM}{\HepChemical{O}{2}{-}\xspace}

\newcommand{\CHFi}{\HepChemical{H}{5}{}\xspace}

\newcommand{\CHoo}{\HepChemical{H}{11}{}}
\newcommand{\CHot}{\HepChemical{H}{12}{}\xspace}
\newcommand{\CHos}{\HepChemical{H}{16}{}\xspace}
\newcommand{\CHoS}{\HepChemical{H}{17}{}\xspace}
\newcommand{\CHtt}{\HepChemical{H}{22}{}\xspace}
\newcommand{\CHGdLS}{\HepChemical{H}{28.128}{}\xspace}
\newcommand{\CHLAB}{\HepChemical{H}{28.28}{}\xspace}
\newcommand{\CHtN}{\HepChemical{H}{2n}{}\xspace}
\newcommand{\CCs}{\HepChemical{C}{6}{}\xspace}
\newcommand{\CCn}{\HepChemical{C}{9}{}\xspace}
\newcommand{\CCoF}{\HepChemical{C}{15}{}\xspace}
\newcommand{\CCGdLS}{\HepChemical{C}{17.072}{}\xspace}
\newcommand{\CCLAB}{\HepChemical{C}{17.14}{}\xspace}
\newcommand{\CCtf}{\HepChemical{C}{24}{}\xspace}
\newcommand{\CCN}{\HepChemical{C}{n}{}\xspace}
\newcommand{\COGdLS}{\HepChemical{O}{0.0126}{}\xspace}
\newcommand{\CNGdLS}{\HepChemical{N}{0.0037}{}\xspace}
\newcommand{\CGdGdLS}{\HepChemical{Gd}{0.0015}{}\xspace}

\newcommand{\CCrt}{\HepChemical{Cr}{2}{}\xspace}
\newcommand{\COT}{\HepChemical{O}{3}{}\xspace}
\newcommand{\CAlt}{\HepChemical{Al}{2}{}\xspace}
\newcommand{\CCsHF}{\CCs\kern-0.2em\CHFi}
\newcommand{\CCnHot}{\CCn\kern-0.2em\CHot}

\newcommand{\CCNHtN}{\CCN\kern-0.2em\CHtN}
\newcommand{\CCHLAB}{\CCLAB\kern-0.2em\CHLAB}

\newcommand{\CCrtOT}{\CCrt\kern-0.25em\COT}

\newcommand{\CAltOT}{\CAlt\kern-0.25em\COT}
\newcommand{\CBaTiOT}{BaTi\kern0.05em\COT}

\newcommand{\CTMHA}{\CCn\kern-0.2em\CHoS\kern-0.2em\COt}
\newcommand{\CTMHAm}{\CCn\kern-0.2em\CHos\kern-0.2em\COt}
\newcommand{\CTMHAM}{\CCn\kern-0.2em\CHos\kern-0.2em\COtM}
\newcommand{\CPPO}{\CCoF\kern-0.2em\CHoo NO}
\newcommand{\CbisMSB}{\CCtf\kern-0.2em\CHtt}
\newcommand{\CGdLS}{\CCGdLS\kern-0.2em\CHGdLS\kern-0.2em\COGdLS\kern-0.2em\CNGdLS\kern-0.2em\CGdGdLS}
\colorlet{mrocol}{ProcessBlue!100!}
\colorlet{lrocol}{SkyBlue!50!}

\newcolumntype{L}[1]{>{\raggedright\arraybackslash}m{#1}}
\newcolumntype{C}[1]{>{\centering\arraybackslash}m{#1}}
\newcolumntype{R}[1]{>{\raggedleft\arraybackslash}m{#1}}

\def\jetset74   {\mbox{\tt Jetset \hspace{-0.5em}7.\hspace{-0.2em}4}\xspace}

\def\ze#1   {\ensuremath{\zeta_{#1}}\xspace}

\def\CO2  {$\mathrm{CO}_2$\xspace}

\usepackage[draft]{pgf}

\begin{document}

\title{Projected WIMP sensitivity of the LUX-ZEPLIN (LZ) dark matter experiment}

\author{D.S.~Akerib}
\affiliation{SLAC National Accelerator Laboratory, Menlo Park, CA 94025-7015, USA}
\affiliation{Kavli Institute for Particle Astrophysics and Cosmology, Stanford University, Stanford, CA  94305-4085 USA}

\author{C.W.~Akerlof}
\affiliation{University of Michigan, Randall Laboratory of Physics, Ann Arbor, MI 48109-1040, USA}

\author{S.K.~Alsum}
\affiliation{University of Wisconsin-Madison, Department of Physics, Madison, WI 53706-1390, USA}

\author{H.M.~Ara\'{u}jo}
\affiliation{Imperial College London, Physics Department, Blackett Laboratory, London SW7 2AZ, UK}

\author{M.~Arthurs}
\affiliation{University of Michigan, Randall Laboratory of Physics, Ann Arbor, MI 48109-1040, USA}

\author{X.~Bai}
\affiliation{South Dakota School of Mines and Technology, Rapid City, SD 57701-3901, USA}

\author{A.J.~Bailey}
\altaffiliation[Now at: ]{University of Valencia, IFC, 46980 Paterna, ESP}
\affiliation{Imperial College London, Physics Department, Blackett Laboratory, London SW7 2AZ, UK}

\author{J.~Balajthy}
\affiliation{University of Maryland, Department of Physics, College Park, MD 20742-4111, USA}

\author{S.~Balashov}
\affiliation{STFC Rutherford Appleton Laboratory (RAL), Didcot, OX11 0QX, UK}

\author{D.~Bauer}
\affiliation{Imperial College London, Physics Department, Blackett Laboratory, London SW7 2AZ, UK}

% ...

\collaboration{The LUX-ZEPLIN Collaboration}

\begin{abstract}
LUX-ZEPLIN (LZ) is a next generation dark matter direct detection experiment that will operate 4850 feet underground at the Sanford Underground Research Facility (SURF) in Lead, South Dakota, USA. Using a two-phase xenon detector with an active mass of 7~tonnes, LZ will search primarily for low-energy interactions with Weakly Interacting Massive Particles (WIMPs), which are hypothesized to make up the dark matter in our galactic halo. In this paper, the projected WIMP sensitivity of LZ is presented based on the latest background estimates and simulations of the detector.
For a 1000~live day run using a 5.6~tonne fiducial mass, LZ is projected to exclude at 90\% confidence level spin-independent WIMP-nucleon cross sections above $1.4 \times 10^{-48}$~cm$^{2}$ for a 40~$\mathrm{GeV}/c^{2}$ mass WIMP. Additionally, a $5\sigma$ discovery potential is projected reaching cross sections below the exclusion limits of recent experiments. For spin-dependent WIMP-neutron(-proton) scattering, a sensitivity of $2.3 \times 10^{-43}$~cm$^{2}$ ($7.1 \times 10^{-42}$~cm$^{2}$) for a 40~$\mathrm{GeV}/c^{2}$ mass WIMP is expected. With underground installation well underway, LZ is on track for commissioning at SURF in 2020.
\end{abstract}

\maketitle


\section{Introduction}

A decade ago, results from the ZEPLIN~\cite{alner:2007ja} and XENON \cite{Angle:2007uj} collaborations ushered in a new era in underground searches for galactic dark matter in the form of Weakly Interacting Massive Particles (WIMPs), dramatically improving the pace at which sensitivity to this candidate particle has progressed. The acceleration was made possible by the introduction of the two-phase (liquid/gas) xenon time projection chamber (LXe-TPC), the origins of which date back to the 1970s~\cite{dolgoshien:1970a,Barabash:1989xr,benetti:1993203}. 

Noble liquid TPCs combine several attractive features for dark matter detectors~\cite{chepel:2012sj}: particle identification to reject backgrounds, 3-D position reconstruction, excellent self-shielding from external backgrounds, and cost-effective scalability compared to solid-state detectors. Liquid xenon (LXe) in particular is an attractive target for WIMP detection due to its efficient conversion of energy from low energy nuclear recoils into observable scintillation and ionization signals. Compared to other noble elements, xenon offers several advantages: an absence of long-lived activation products; high sensitivity to spin-independent (SI) WIMP interactions due to its large atomic mass and a coherent scattering enhancement ($\propto A^{2}$) for non-relativistic WIMPs, assuming isospin-conserving interactions; and sensitivity to spin-dependent (SD) interactions due to naturally-occurring odd-neutron isotopes. Having probed SI cross sections as low as $10^{-47}$~cm$^2$~\cite{Akerib:2016vxi, Cui:2017nnn, Aprile:2018dbl}, LXe-TPCs are leading the search for WIMP dark matter above a few GeV/$c^2$ mass.

Formed by the merger of the LUX and ZEPLIN-III collaborations, LUX-ZEPLIN (LZ) is constructing a next generation dark matter detector using a LXe-TPC with an expected SI(SD) sensitivity in the low $10^{-48}(10^{-43})$~cm$^{2}$ range. To achieve this, LZ overcomes key experimental challenges: a powerful active veto system and a comprehensive radio-assay and surface cleanliness program ensure an ultra-low background environment; and a 7~tonne active mass provides both self-shielding and sufficient target mass for a 15~$\mathrm{tonne} \cdot \mathrm{year}$ exposure, while at the same time maintaining the light and charge collection in the TPC necessary to detect low-energy nuclear recoils. 

This paper presents the expected WIMP sensitivity of the experiment, along with the main components in deriving this sensitivity: detector design and parameters, background simulations, material assay results, and the statistical procedure for WIMP sensitivity analysis.

\section{The LZ Instrument}

\subsection{Overview}

\begin{figure*}[t!]
\begin{tabular}{ccc}
\includegraphics[trim={0 0.0cm 0 0.0cm},clip,width=0.44\linewidth]{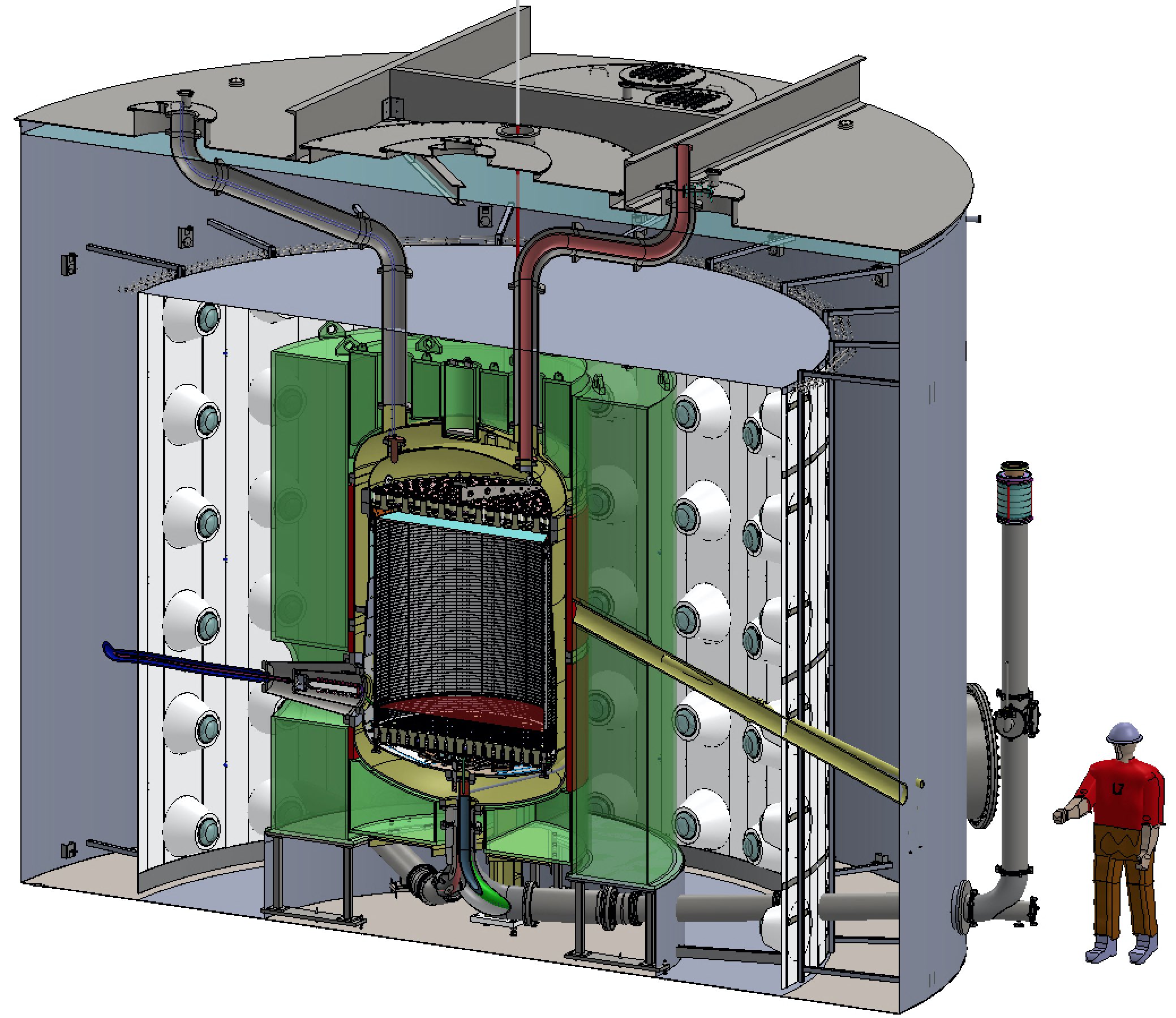} & &
\includegraphics[trim={0 0.0cm 0 0.2cm},clip,width=0.37\linewidth]{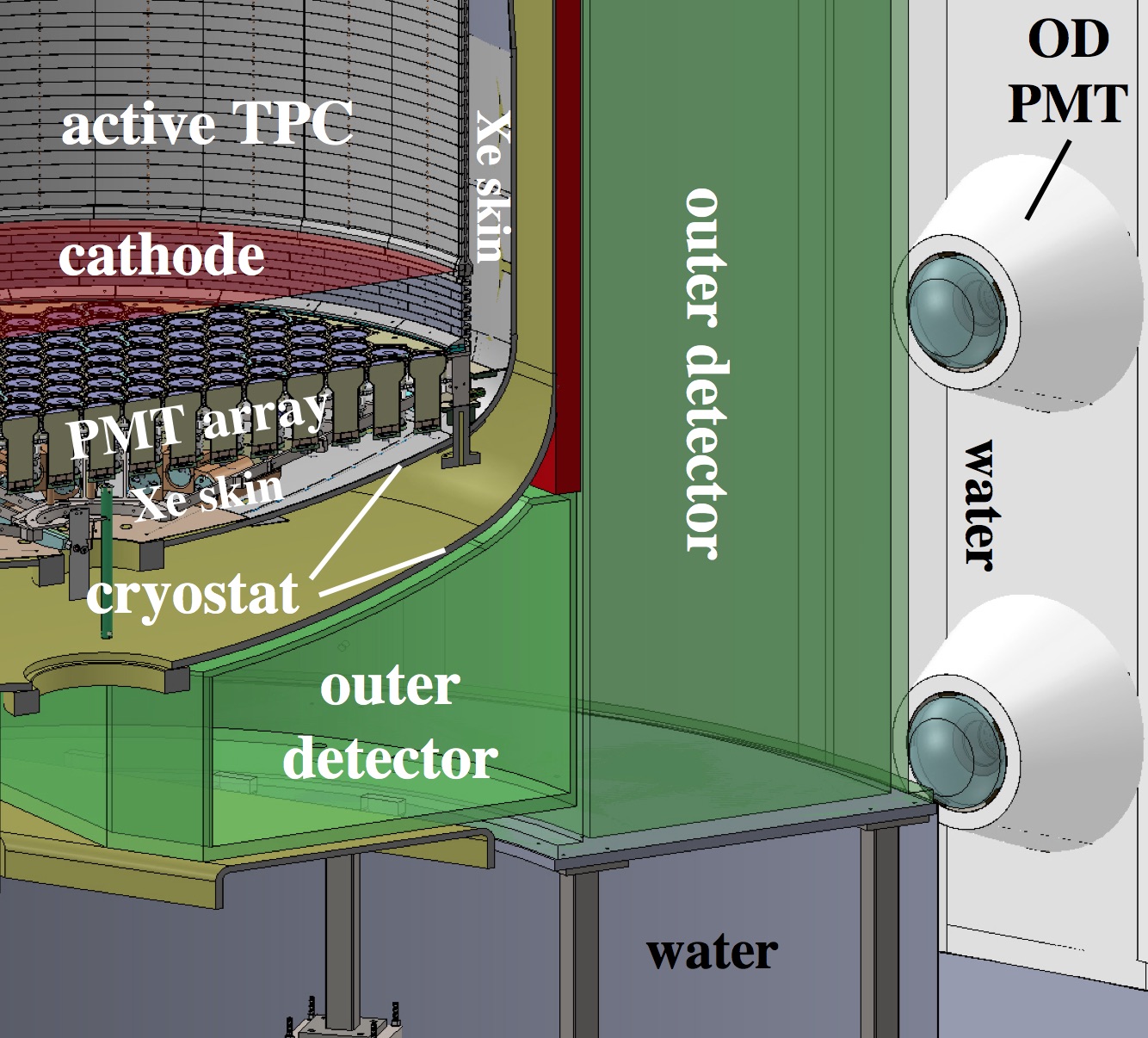} \\
\end{tabular}
\caption{\label{LZSolid} Left: Cutaway drawing of the LZ detector system. The LXe-TPC is surrounded by the outer detector (OD) tanks (green) and light collection system (white), all housed in a large water tank (blue-grey). Conduits penetrate the various regions and boundaries to deliver services to the LXe-TPC: PMT and instrumentation cables (top and bottom, red); cathode high voltage (lower left with cone); purified LXe (bottom center, green); neutron beam conduit (right, yellow and pitched). 
Right: Expanded view of the lower right corner. `OD PMT' indicates an outer detector photomultiplier tube. The xenon skin region is observed by an independent set of PMTs (not depicted).}
\end{figure*}

A cutaway drawing of the experiment is shown in Fig.~\ref{LZSolid}. The vacuum-insulated cryostat made from ultra-pure titanium~\cite{Akerib:2017iwt} holds 10 tonnes of LXe, including the LXe-TPC and its enveloping xenon skin veto. The cryostat is maintained at 175~K by a system of thermosyphons and is surrounded by a room temperature liquid scintillator outer detector (OD). Both are located within a large water tank in the Davis Campus at the \num{4850}-foot level (4300~m~w.e.) of the Sanford Underground Research Facility (SURF)~\cite{Heise:2017rpu}. Key dimensions and masses of the experiment are summarized in Table~\ref{table:dimensions}.

The active volume of the TPC is a cylinder with both diameter and height equal to 1.46~m, containing 7-tonnes of LXe. Particle interactions in the LXe generate prompt scintillation light (`S1') and release ionization electrons---the latter drift in an applied vertical ($z$) electric field and are extracted into the gas layer above the surface where they generate electroluminescence photons (`S2'). The xenon circulation and purification strategies are based on the LUX experience~\cite{akerib:2012ya,akerib:2012ys,akerib:2016hcd} and electronegative impurities are suppressed sufficiently to allow electrons to survive, with good efficiency, drifting through the length of the TPC.

Photons are detected by 494 Hamamatsu R11410-22 3$^{\prime\prime}$-diameter photomultiplier tubes (PMTs), with a demonstrated low level of radioactive contamination~\cite{akerib:2012da,Aprile:2015lha} and high quantum efficiency~\cite{Paredes:2018hxp} at the LXe scintillation wavelength of 175~nm~\cite{Fujii:2015293}. The PMTs are assembled in two arrays viewing the LXe from above and below. The 241 bottom PMTs are arranged in a close-packed hexagonal pattern to maximize the collection efficiency for S1 light. The 253 top PMTs are arranged in a hybrid pattern that transitions from hexagonal near the center to nearly circular at the perimeter, thereby optimizing the $(x,y)$ position reconstruction of the S2 signal for interactions near the TPC walls. The TPC walls are made of highly reflective polytetrafluoroethylene (PTFE) panels that also embed \num{57} field-shaping rings which define the drift field.

Vertical electric fields in the TPC are created by four horizontal electrode planes, which consist of grids woven from thin stainless steel wires. At the top of the TPC, the gate and anode grids (operating at $\mp 5.75$~kV, respectively) straddle the liquid surface to extract ionization electrons from the liquid into the gas, and to create an S2-generating region in the gas phase. At the bottom, the cathode grid defines the lower boundary of the active TPC volume. An additional grid below the cathode shields the bottom PMT array from the cathode potential. This creates a reverse field region below the cathode, containing 840~kg of LXe, where energy deposits create S1-only events. The drift field is established between the cathode and gate grid. The nominal cathode operating voltage is $-50$~kV, delivered from a dedicated conduit penetrating the cryostat laterally. In this work we assume a uniform TPC drift field of 310~\si{\volt\per\centi\meter}. 

\begin{table}[t!]
\centering
\caption{\label{table:dimensions} Summary of key dimensions and masses. The inner cryostat and the xenon skin region have a tapered radial profile as indicated. Top and bottom liquid scintillator (GdLS) tanks also have a range of dimensions. The xenon skin thickness below the lower PMT array is omitted in this table due to the complexity of the geometry.}
\smallskip
\begin{tabular}{ m{2cm}  l | c }   
\hline \hline
\multicolumn{2}{l}{Parameter [units]} & Value \\ \hline \hline
\multicolumn{2}{l}{\quad TPC active height [\si{\meter}] } & 1.46 \\ 
\multicolumn{2}{l}{\quad TPC inner diameter [\si{\meter}] } & 1.46 \\ 
\multicolumn{2}{l}{\quad active LXe mass [\si{\kg}] } & 7000 \\ 
\multicolumn{2}{l}{\quad xenon skin thickness, side [\si{\centi\meter}] } & 4.0 - 8.0 \\ 
\multicolumn{2}{l}{\quad inner cryostat diameter  [\si{\meter}] } & 1.58 - 1.66 \\
\multicolumn{2}{l}{\quad inner cryostat height [\si{\meter}]} & 2.59 \\
\multicolumn{2}{l}{\quad outer cryostat inside diameter [\si{\meter}] } & 1.83 \\
\multicolumn{2}{l}{\quad outer cryostat height [\si{\meter}] } & 3.04 \\
\multicolumn{2}{l}{\quad GdLS tanks outer radius [\si{\meter}] } & 1.64 \\
\multicolumn{2}{l}{\quad GdLS thickness, side [\si{\centi\meter}] } & 61 \\
\multicolumn{2}{l}{\quad GdLS thickness, top [\si{\centi\meter}] } & 40 - 62 \\
\multicolumn{2}{l}{\quad GdLS thickness, bottom [\si{\centi\meter}] } & 34.5 - 57 \\
\multicolumn{2}{l}{\quad GdLS mass [\si{\ton}] } & 17.3 \\
\multicolumn{2}{l}{\quad GdLS tanks, acrylic wall thickness [\si{\centi\meter}] } & 2.54 \\
\multicolumn{2}{l}{\quad water thickness, GdLS vessels to PMTs [\si{\centi\meter}] } & 84 \\
\multicolumn{2}{l}{\quad water tank diameter [\si{\meter}] } & 7.62 \\ 
\multicolumn{2}{l}{\quad water tank height [\si{\meter}] } & 5.92 \\ 
\multicolumn{2}{l}{\quad water mass [\si{\ton}] } & 228 \\ \hline \hline
\end{tabular}
\end{table}

A two-component veto system rejects multi-site backgrounds and asynchronously characterizes the radiation environment around the WIMP target. The innermost veto component is the xenon skin region, formed by instrumenting the outer 2~tonnes of LXe located between the TPC and the inner cryostat vessel. This region is optically segregated from the TPC, and scintillation light produced in the LXe is viewed by 93~Hamamatsu R8520 1$^{\prime\prime}$ PMTs mounted near the xenon liquid level and a further 38 Hamamatsu R8778 2$^{\prime\prime}$ PMTs mounted near the bottom of the TPC. The inner surface of the inner cryostat vessel is covered by a thin liner of PTFE to improve light collection. The principal role of this skin region is the detection of scattered gamma rays. A 3~phd requirement on the scintillation signal yields an effective analysis threshold of 100~keV for more than 95\% of the skin volume.

The second veto component is the OD that surrounds the LZ cryostat. It constitutes a near-hermetic layer, formed by 17~tonnes of gadolinium-loaded liquid scintillator (GdLS)~\cite{Yeh:2007zz,Ding:2008zzb}, contained in 10 acrylic tanks. The principal role of the OD is the tagging of neutrons which emerge after causing nuclear recoils in the TPC, a background otherwise indistinguishable from WIMP recoils on an event-by-event basis. Neutrons tend to scatter off high-Z components of LZ until they diffuse to and moderate on the hydrogen in the GdLS, after which they capture on the gadolinium, releasing approximately 8~MeV in a cascade of gamma rays (average multiplicity 4.7). The capture follows an approximate exponential time distribution after either an S1 signal in the LXe-TPC or a prompt proton-recoil signal in the OD with a time constant of 28~\si{\micro\second}~\cite{An:2016ses}. Some neutrons diffuse out of, then return to the GdLS resulting in a capture time constant closer to 200~\si{\micro\second}, necessitating a benchmark time window of 500~\si{\micro\second} for efficient tagging. Scintillation light produced in the GdLS is observed by 120 Hamamatsu R5912 8$^{\prime\prime}$ PMTs mounted in the water space outside of the acrylic tanks and surrounded by Tyvek diffuse reflectors. The volume outside the acrylic tanks is filled with ultrapure water, providing suppression of backgrounds from naturally occurring radiation in the surrounding rock in the Davis Campus, and from the OD PMTs. The OD light collection system yields an effective energy threshold of 100~keV. To reduce dead time from the radioactive decays of \ICof, \ISmofS and \IGdoFt, an analysis threshold of 200~keV is assumed, providing a greater than 95\% veto efficiency for neutrons that scatter once in the TPC. The dead time induced by all sources internal and external to the GdLS is lower than 5\%.

A delivery system for sealed radioactive sources to the vacuum space between the two cryostat vessels and an injection system for dispersible radioisotopes into the xenon flow allow the TPC, the xenon skin, and OD to be calibrated with a suite of beta, gamma, and neutron sources. An external deuterium-deuterium neutron generator is employed outside of the water tank, with air-filled conduits (visible in Fig.~\ref{LZSolid}) providing collimation of the neutron beam~\cite{Akerib:2016mzi}. Photoneutron sources are deployed through a guide tube and sit on top of the cryostat, providing a source of mono-energetic neutrons with nuclear recoil end points below 5~keV.   

Key technical challenges have been addressed during the design and construction of the instrument, including: TPC high voltage performance, PMT characterization and quality assurance, measurements of PTFE optical properties and demonstrations of the main calibration systems. A wide array of system tests is in place to ensure that all detector requirements are adequately met. A comprehensive account of all aspects of the experiment can be found in the LZ Technical Design Report (TDR)~\cite{Mount:2017qzi}.

\subsection{Experimental strategy}

The xenon target material is monitored for evidence of excess nuclear recoils that may be attributed to WIMP dark matter scattering, in particular for single scattering events occurring in an inner 5.6 tonne fiducial volume and within an energy range of interest relevant for WIMPs.
Primary backgrounds are of the electronic recoil (ER)
and nuclear recoil (NR) varieties. Most background events are 
due to intrinsic radioactivity in the xenon, the detector
materials, and the experimental hall. Many gamma and neutron events can be rejected
by requiring that no energy is observed in the xenon skin and the OD. 
The remaining set of WIMP candidates are examined with a Profile Likelihood Ratio (PLR) fit. At present the fit is performed in a two-dimensional (S1,S2) space, and it distinguishes between NR and ER events due to their differing relative yields of scintillation and ionization; ultimately the fit will be extended to include position information with the aim to utilize an expanded LXe volume. WIMP signal distributions are simulated for a variety of WIMP mass hypotheses and each is tested for its compatibility with the data. As detailed in Sec.~\ref{sec:BGs}, the most important backgrounds to the WIMP signal are beta decays in the LXe (mostly radon daughter species such as $^{214}$Pb and $^{212}$Pb, as well as $^{85}$Kr), ER events from $pp$ solar neutrinos scattering with atomic electrons, and NR events from coherent scattering of atmospheric neutrinos. Coherent nuclear scattering of $^{8}$B solar neutrinos is an important source of very low energy NR events.

\subsection{Key experimental parameters} 

Table~\ref{table:parameters} lists the key detector parameters for the LXe-TPC, based on measurements made of actual materials and components procured for use in LZ. More conservative baseline parameters were described in the TDR~\cite{Mount:2017qzi} and represent the minimum requirements that have been set for basic functionality. The photon detection efficiency in liquid, $g_1$, is the average fraction of S1 light produced in the TPC that is eventually detected by any of the 494 TPC PMTs; $g_{1,\mathrm{gas}}$ is the equivalent detection efficiency for S2 electroluminescence photons generated in the extraction region. S1 and S2 signals are measured in units of `photons detected' (phd), an observable that accounts for double photoelectron emission from the PMT photocathode at these wavelengths~\cite{Faham:2015kqa,Akerib:2017vbi}. The current estimate of $g_1$ is $11.9\%$, both it and $g_{1,\mathrm{gas}}$ are derived from optical simulations based on reflectivity measurements of the LZ PTFE~\cite{silva:2009ip,Haefner:2016ncn,neves:2016tcw}; measurements of the quantum efficiency, first dynode collection efficiency, and two photoelectron emission probability in a sample of the 3$^{\prime\prime}$ Hamamatsu PMTs to be used in LZ~\cite{Paredes:2018hxp}; and a photon absorption length in LXe motivated by the high light yields reported in the literature~\cite{Akerib:2016qlr,Aprile:2017aty}. The electron extraction efficiency, not included in $g_{1,\mathrm{gas}}$, is extrapolated from~\cite{russians1979}. Finally, the trigger efficiency for single photoelectrons (phe) is based on measurements of a full-scale LZ electronics test chain described in the LZ TDR. The S1 coincidence level, electron extraction efficiency, drift field and electron lifetime are unchanged from their TDR baseline values.

\begin{table}[t]
\centering
\caption{\label{table:parameters} Key detector parameters for the LXe-TPC. Entries that are offset indicate quantities derived from preceding parameters.}
\smallskip
\begin{tabular}{ m{2cm}  l | r }   
\hline
\multicolumn{2}{c}{Detector Parameter} & Value \\ \hline \hline

\multicolumn{2}{l}{PTFE-LXe(GXe) reflectivity} & 0.977(0.85) \\
\multicolumn{2}{l}{LXe(GXe) photon absorption length [m]} & 100(500) \\
\multicolumn{2}{l}{PMT efficiency$^{*}$ at 175~nm} & 0.27 \\
\multicolumn{2}{l}{\quad\quad Average PDE in liquid ($g_1$) [phd/ph]} & 0.12 \\
\multicolumn{2}{l}{\quad\quad Average PDE in gas ($g_{1,\mathrm{gas}}$) [phd/ph]} & 0.10 \\ \hline
\multicolumn{2}{l}{Single electron size [phd]} & 83 \\
\multicolumn{2}{l}{S2 electron extraction efficiency} & 0.95 \\
\multicolumn{2}{l}{\quad\quad Effective charge gain ($g_2$) [phd/e]} & 79 \\ \hline
\multicolumn{2}{l}{Single phe trigger efficiency} & 0.95 \\
\multicolumn{2}{l}{Single phe relative width (Gaussian)} & 0.38 \\
\multicolumn{2}{l}{S1 coincidence level} & 3-fold \\ \hline
\multicolumn{2}{l}{Drift field [\si{\volt\per\centi\meter}] } & 310 \\
\multicolumn{2}{l}{Electron lifetime [\si{\micro\second}] } & 850 \\ \hline
\multicolumn{2}{l}{{\footnotesize* Including first dynode collection efficiency.}}

\end{tabular}
\end{table}

\section{Simulations}
\label{sec:Sims}

\subsection{Simulations framework}
\label{sec:sims}
A variety of software packages is employed to simulate the physics of
signals and backgrounds that induce responses in the LXe-TPC, xenon skin and OD. The overall 
simulation framework is BACCARAT, which is 
based on an earlier simulation package developed for the LUX
experiment~\cite{Akerib:2011ec}. BACCARAT is built on the GEANT4
toolkit~\cite{Agostinelli:2002hh} and provides object-oriented design
specifically tuned for noble liquid detectors; it records particle
interactions on a geometry-component basis, but with an infrastructure
which is independent of the actual detector geometry.

The results described in this paper were produced with GEANT4 version
9.5 compiled with CLHEP version 2.1.0.1 libraries. Standard GEANT4
optical processes were used to evaluate the $g_1$ and
$g_{1,\mathrm{gas}}$ parameters, but final background analyses were
performed with NEST (Noble Element Simulation Technique) as described in Sec.~\ref{sec:nest}. For
electromagnetic processes the Livermore physics list was used, with
the addition of the Goudsmit-Saunderson Model for multiple
scattering. The hadronics physics list is based on QGSP\_BIC\_HP.
The effects of proton molecular binding on neutron transport and
capture, described by the thermal elastic scattering matrix
$S(\alpha,\beta)$, were not considered.

\begin{figure*}[t]
\centering
  \includegraphics[trim={0 0.15cm 0 0.17cm},clip,width=0.88\linewidth]{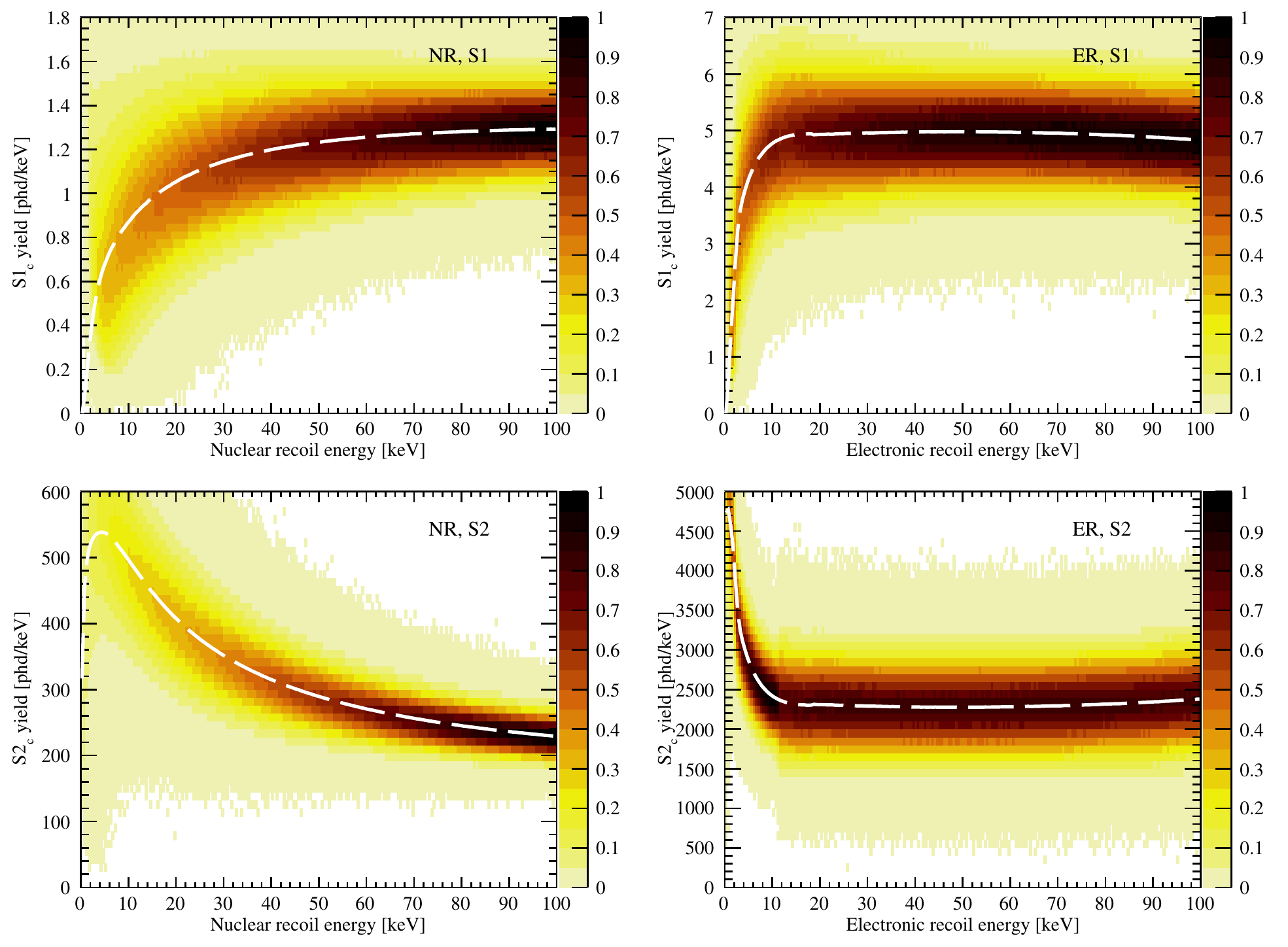}
\caption{Distribution of S1 (top) and S2 (bottom) yields as a function of deposited energy for nuclear (left) and electronic recoils (right) in LZ. The dashed line indicates the average response.}
\label{fig:NEST_response_small}
\end{figure*}

The capability of the simulations framework has been enhanced to address various phenomena that GEANT4 does not model adequately for LZ. An improved description of the de-excitation cascade following neutron capture on \IGd was implemented with use of the DICEBOX algorithm~\cite{Becvar:1998}. The GEANT4 deficiencies in \Pgg emission after neutron capture on other nuclei more complex than the proton were not corrected. A custom event generator was developed to simulate neutron production in materials from naturally occurring uranium and thorium chains using SOURCES-4A~\cite{Wilson:1999a}, modified as described in~\cite{Tomasello:2010zz} (see also references therein). The emission of gamma rays in coincidence with ($\alpha,n$) reactions was not simulated.

A new spontaneous fission (SF) generator was written, implementing particle multiplicity such that the rejection efficiency for decays producing multiple neutrons and gamma rays can be correctly determined. The MUSUN muon simulation code~\cite{Kudryavtsev:2008qh} has been integrated into BACCARAT as a particle generator to sample atmospheric muons around the underground laboratory for further tracking by GEANT4. Finally, a new radioactive decay generator for gamma ray emission from naturally occurring uranium and thorium decay chains was implemented, which allows splitting the activity by individual isotope during analysis, simplifying the implementation of breaks in secular equilibrium.

\subsection{From energy deposition to signals}
\label{sec:nest}

The NEST~\cite{Szydagis:2011tk,Szydagis:2013sih} package is used to convert GEANT4 energy deposits in the xenon skin and TPC volumes to primary scintillation photons and ionisation electrons. Energy deposits are categorised as either ER or NR and separate deposits are summed together assuming a 400~\si{\micro\meter} NEST track cluster size. The NEST model used in this simulation has been updated to incorporate the latest calibration results from the LUX experiment~\cite{Akerib:2015rjg,Akerib:2015wdi,Akerib:2016mzi,Akerib:2016qlr}. For NR energies below 1.1~keV, the lowest energy for which light yield was measured in~\cite{Akerib:2016mzi}, the signal yields are extrapolated down to 0.1~keV following the Lindhard model~\cite{lindhard:1961zz,lindhard:1963}. The impact of this is discussed in Sec.~\ref{sec:WIMPsens}.

The S1 photons are propagated to the faces of the PMTs using GEANT4 with standard optical processes. An average PMT response that includes binomial fluctuations is applied and the response from all PMTs is summed to give the S1 signal size; for these projections a full waveform simulation is not performed. A similar treatment is applied to the S2 light, after the ionization electrons are converted to a number of electroluminescence photons with NEST. S1 is corrected for the variation of light collection with position in the detector (denoted S1$_c$) to match the average S1 response of the active region. S2 is also corrected for the position of the event, including the effect of finite electron lifetime (the mean time an electron remains in the LXe before being attached to an impurity), assumed to be \SI{850}{\micro\second}. Longer electron lifetimes than this have been demonstrated in LUX~\cite{Akerib:2017vbi}. For an electron drift speed of 1.8~\si{\milli\meter\per\micro\second}, expected for the nominal drift field of 310~\si{\volt\per\centi\meter}~\cite{Albert:2016bhh}, this corresponds to a charge absorption length of \SI{1.5}{\meter}.
The corrected signal is denoted S2$_c$, and it is normalized to the average response for interactions uniform in the horizontal plane just below the gate grid. Figure~\ref{fig:NEST_response_small} shows the corrected S1 and S2 yields as a function of deposited energy for both nuclear and electronic recoils in LZ. 

When running the PLR analysis a parametrization of both the NEST and detector response, based on full optical simulations using BACCARAT, allows for fast generation of the (S1,S2) probability density functions (PDFs), with the option to change the detector parameters ($g_1$, $g_2$, S1 coincidence level, drift field, etc.) at runtime. Statistical fluctuations and charge and light $\mathrm{(anti\text{-})correlations}$ are accounted for.

\subsection{Analysis cuts}
\label{sec:analysiscuts}

\begin{figure*}[t!]
\begin{tabular}{ccc}
\includegraphics[trim={0 0.2cm 0 1.3cm},clip,width=0.44\linewidth]{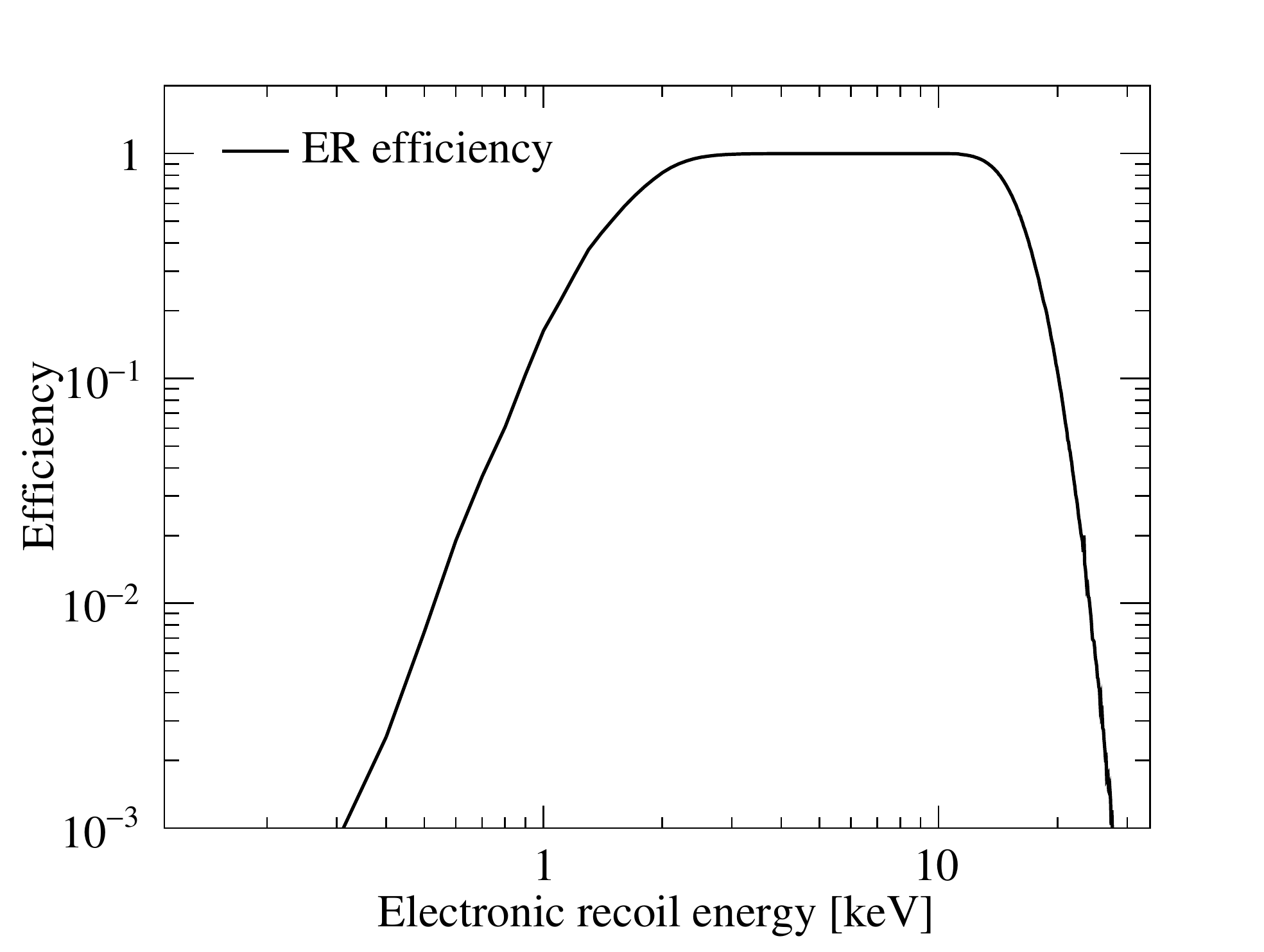} & &
\includegraphics[trim={0 0.2cm 0 1.3cm},clip,width=0.44\linewidth]{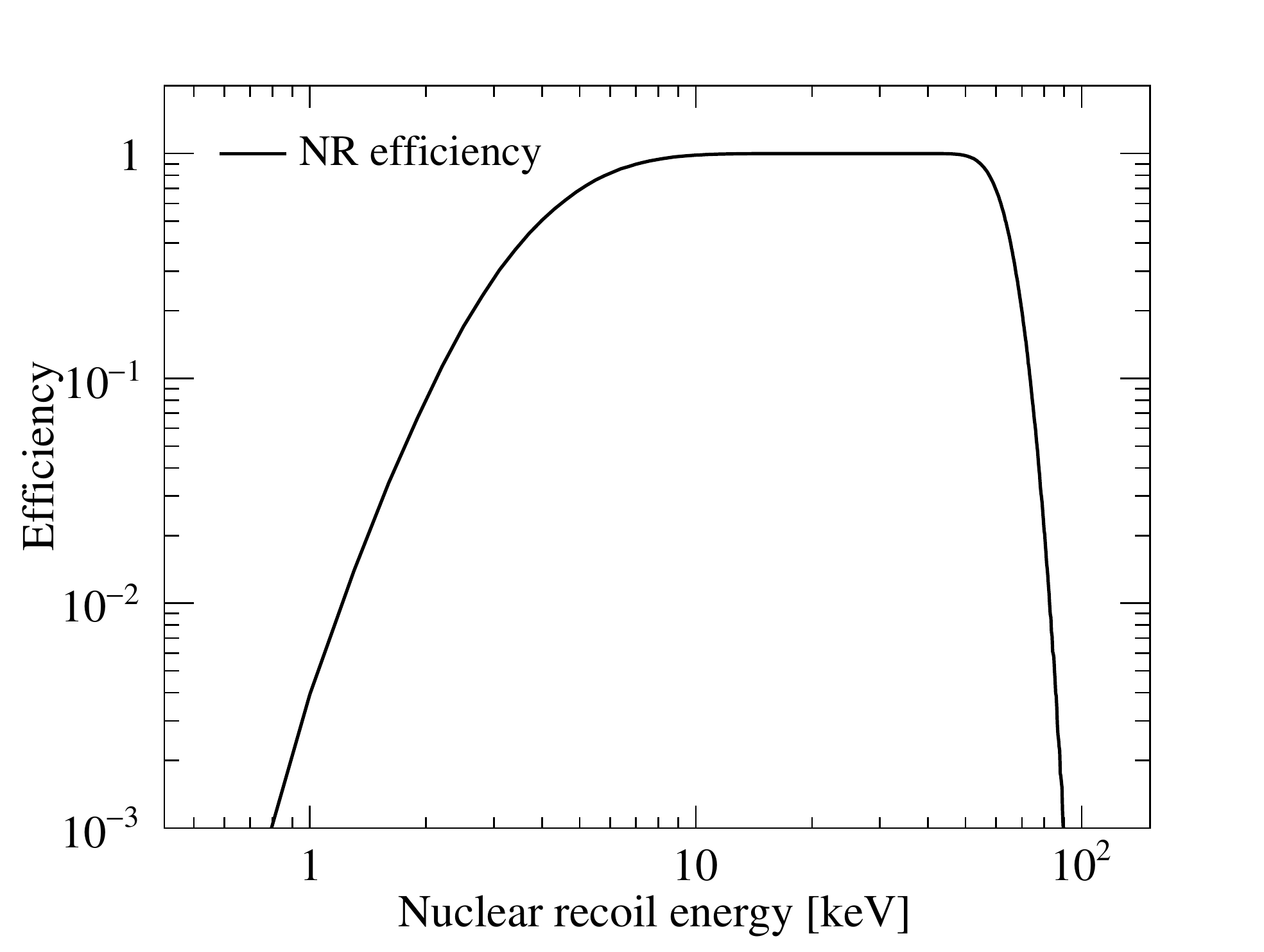} \\
\end{tabular}
\caption{Simulated efficiencies for electronic (left) and nuclear recoils (right) after WIMP search region of interest cuts: 3-fold S1 coincidence, S$2>415$~phd (5 emitted electrons), and S1$_c<80$~phd.}
\label{fig:nr_er_efficiency}
\end{figure*}

A set of cuts is applied to the simulated data to select WIMP-like events and determine the impact of backgrounds on the WIMP-search analysis.

A single scatter (SS) cut rejects multiple-scattering neutrons and gammas by requiring that the energy-weighted standard deviation ($\sigma_z$ and $\sigma_r$) of any separate NEST clusters is less than the expected spatial resolution of the detector. Based on the LUX position reconstruction~\cite{Akerib:2017vbi} and accounting for the larger size and separation of LZ PMTs a conservative requirement is $\sigma_z<0.2$~cm and $\sigma_r<3.0$~cm for S2 signals at the detection threshold. For actual data analysis this cut will be replaced with analysis routines based on hit patterns and waveform shapes. Next, events outside the WIMP search region of interest (ROI) are removed: the S1 signal must have at least 3-fold coincidence in the TPC PMTs and have a total corrected S1$_{c}$ size of less than 80~phd. In addition, the uncorrected S2 signal is required to be greater than 415~phd (5~emitted electrons), ensuring adequate signal size for position reconstruction.

TPC events with a time-coincident signal in either of the veto detectors are removed: for the xenon skin at least 3~phd must be observed within an \SI{800}{\micro\second} coincidence window before or after the time of the TPC S1 signal, whilst for the OD at least 200~keV must be deposited within \SI{500}{\micro\second}. These time intervals ensure vetoing both prompt gammas and the delayed signals from thermal neutron capture.

Lastly, a fiducial volume (FV) cut removes background events near the edges of the TPC. The FV is cylindrical with boundaries defined to be 4~cm from the TPC walls, 2~cm above the cathode grid (with 14.8~cm of LXe below the cathode providing further shielding) and 13~cm below the gate grid. The fiducial volume contains 5.6~tonnes of LXe. The misreconstruction of wall events into the fiducial volume drives the choice of a mostly cylindrical volume: studies of position reconstruction of simulated S2 edge events, using the Mercury algorithm~\cite{solovov:2011aa,Akerib:2017riv}, indicate that this probability falls sharply as a function of distance to wall. At 4~cm it is less than $10^{-6}$ for the smallest S2 signals considered, ensuring that wall events are a sub-dominant background. Ultimately, inclusion of spatial coordinates in the PLR will obviate the need for a fixed fiducial volume.

Figure~\ref{fig:nr_er_efficiency} shows the simulated efficiencies after application of the WIMP search ROI cut for single scatter events in the TPC as a function of recoil energy for electronic and nuclear recoils. This region of interest specifically targets SI and SD WIMP recoils ($\lesssim 100$~keV). Searches for other physics signals such as dark matter interacting through non-relativistic effective field theory operators~\cite{fan:2010gt,fitzpatrick:2013lia}, inelastic dark matter~\cite{TuckerSmith:2001hy,Akimov:2010vk,baudis:2013bba}, and neutrinoless double-beta decay~\cite{ndbd:2015} will focus on different energy ranges.

These cuts are applied separately to simulations of each background source in each detector component (around 200 component-source pairs in total) to obtain a probability of a background event being identified as a candidate WIMP event. These are then combined with material activities from the radio-assay program described in Sec.~\ref{sec:BGs} to estimate the rate at which background events are misidentified as WIMP candidates in LZ.

\section{Backgrounds}
\label{sec:BGs}

Measured material radioactivity and anticipated levels of dispersed and surface radioactivity are combined with the Monte Carlo simulations and analysis cuts described in Sec.~\ref{sec:Sims} to determine background rates in the detector. Table~\ref{table:backgrounds} presents integrated background ER and NR counts in the 5.6~tonne fiducial mass for a 1000 live day run using a reference cut-and-count analysis, both before and after ER discrimination cuts are applied.
For the purposes of tracking material radioactivity throughout the design and construction of LZ, the counts in Table~\ref{table:backgrounds} do not use the ROI described in Sec.~\ref{sec:Sims} and instead are for a restricted region relevant to a 40~GeV/c$^{2}$ WIMP spectrum, equivalent to approximately 1.5--6.5~keV for ERs and 6--30~keV for NRs. For continuity with previous studies the values in Table~\ref{table:backgrounds} are based on the baseline optical model described in the TDR~\cite{Mount:2017qzi}; when constructing the background model used for the sensitivity projections in Sec.~\ref{sec:WIMPsens} the full ROI and the projected optical model described in Table~\ref{table:parameters} are used.

\begin{figure*}[t!]
\begin{tabular}{ccc}
\includegraphics[trim={0 0.15cm 0.5cm 1.8cm},clip,width=0.44\linewidth]{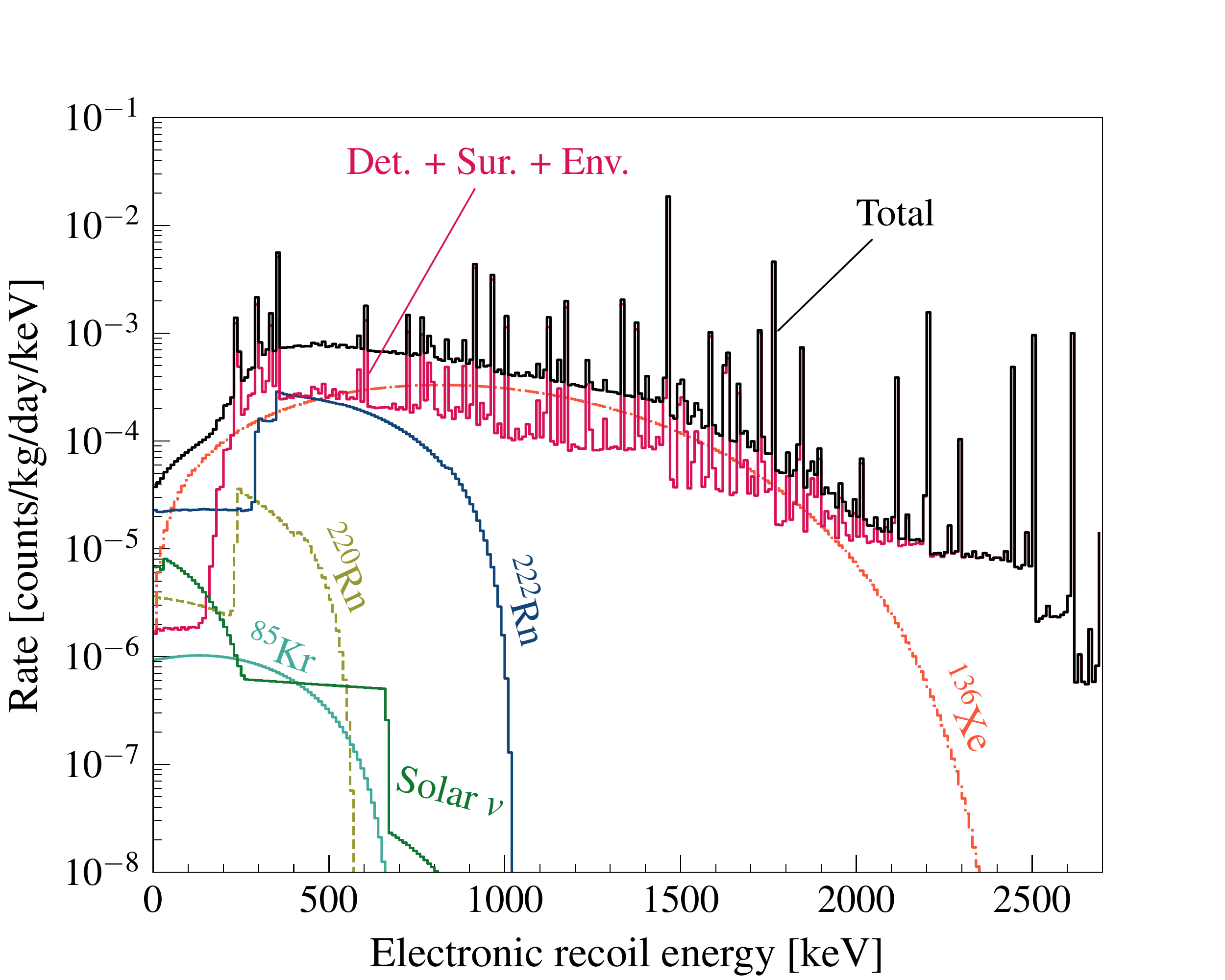} & &
\includegraphics[trim={0 0.15cm 0.5cm 1.8cm},clip,width=0.44\linewidth]{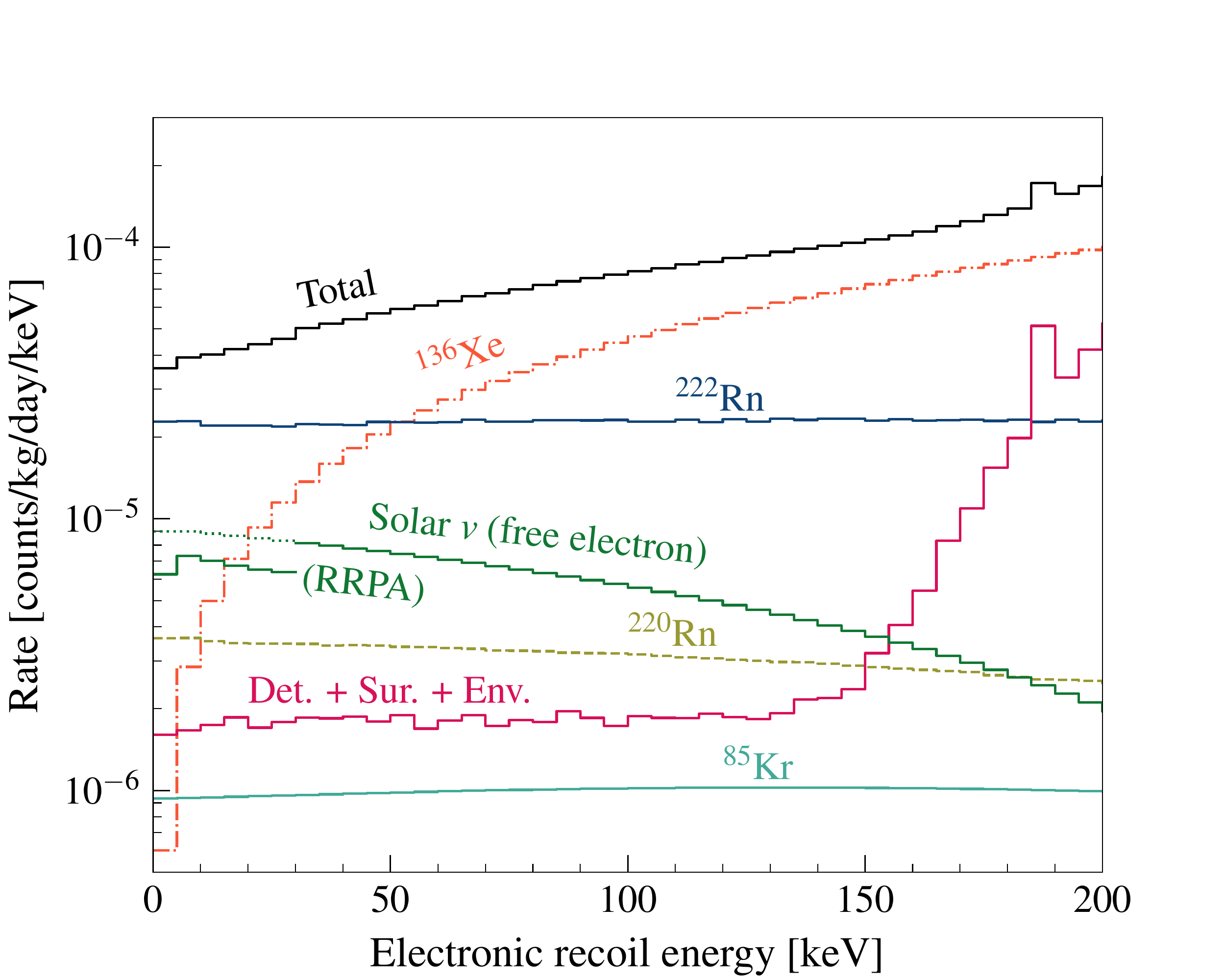} \\
\end{tabular}
\caption{ER background spectra in the 5.6-tonne fiducial volume for single scatter events with neither a xenon skin nor an OD veto signal. No detector efficiency or WIMP-search region of interest cuts on S1$_c$ have been applied. The right-hand panel shows a close-up of the 0--200~keV region of the left-hand panel. Below 30~keV the contribution from elastic scattering of $pp$+$^{7}$Be+$^{13}$N solar neutrinos is scaled according to the relativistic random phase approximation (RRPA) calculation in~\cite{Chen:2017eab}.}
\label{fig:DRU_ER}
\end{figure*}

\begin{figure}[t!]
\centering
\includegraphics[trim={0 0.5cm 1.1cm 2.0cm},clip,width=1.0\linewidth]{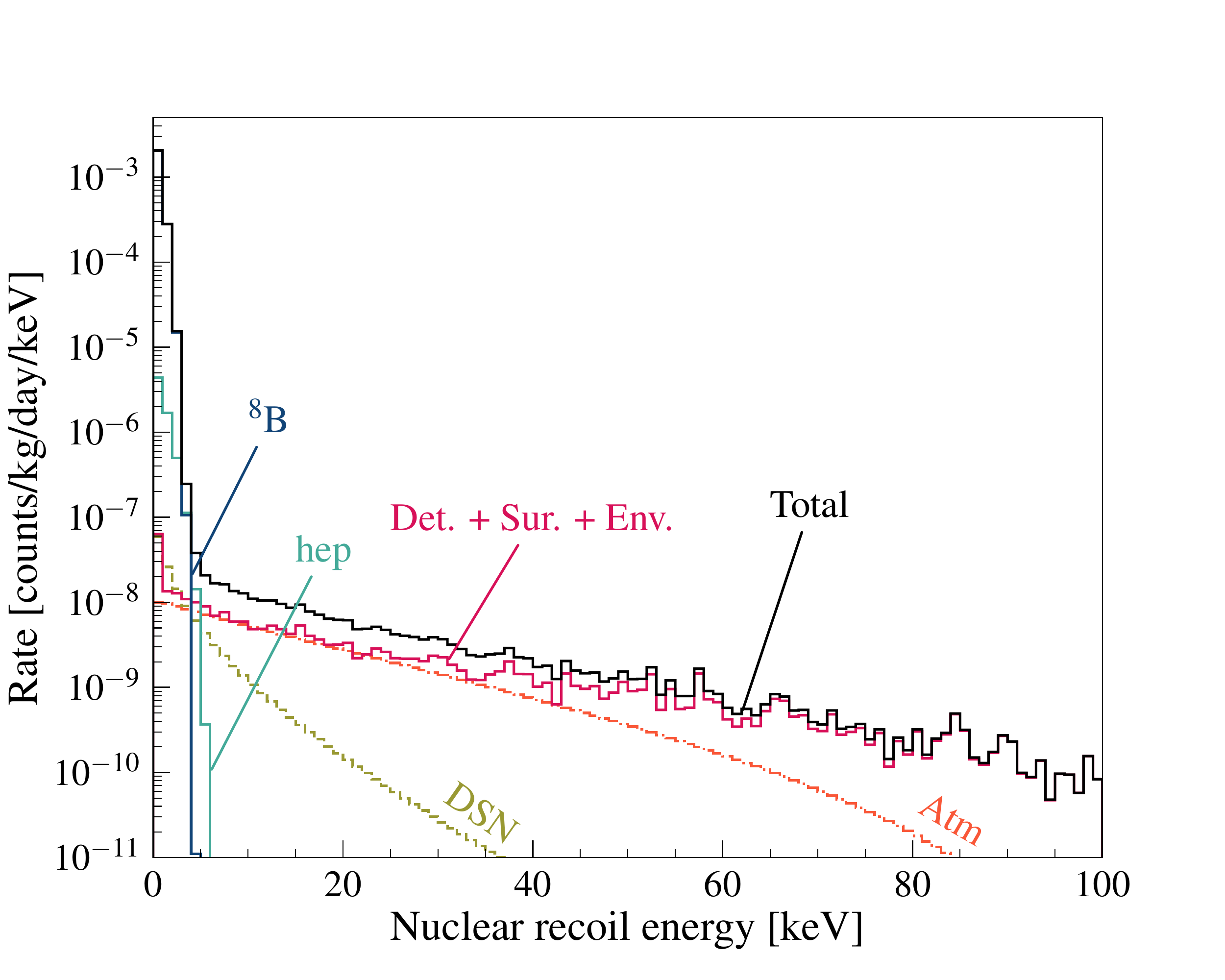}
\caption{NR background spectra in the 5.6-tonne fiducial volume for single scatter events with neither a xenon skin nor an OD veto signal. No detector efficiency or WIMP-search region of interest cuts on S1c have been applied.}
\label{fig:DRU_NR}  
\end{figure}

The expected total from all ER(NR) background sources is 1131(1.03) counts in the full 1000~live day exposure. Applying discrimination against ER at 99.5\% for an NR acceptance of 50\% (met for all WIMP masses given the nominal drift field and light collection efficiency in LZ~\cite{Mount:2017qzi}) suppresses the ER(NR) background to 5.66(0.52) counts. Radon presents the largest contribution to the total number of events. Atmospheric neutrinos are the largest contributor to NR counts, showing that LZ is approaching the irreducible neutrino background~\cite{billard:2013qya}. Figures~\ref{fig:DRU_ER} and~\ref{fig:DRU_NR} show the spectral contributions to ER and NR backgrounds, respectively, used when generating the (S1,S2) PDFs for the sensitivity analysis described in Sec.~\ref{sec:WIMPsens}. These figures show rates of un-vetoed single scatter events in the fiducial volume with no energy region of interest or detector efficiency cuts applied. 

\begin{table*}[ht!]
\centering
\caption[BackgroundsTable]{Estimated backgrounds from all significant sources in the LZ 1000~day WIMP search exposure. Counts are for a region of interest relevant to a 40~GeV/c$^{2}$ WIMP: approximately 1.5--6.5~keV for ERs and 6--30~keV for NRs; and after application of the single scatter, skin and OD veto, and 5.6~tonne fiducial volume cuts. Mass-weighted average activities are shown for composite materials and the $^{238}$U and $^{232}$Th chains are split into contributions from early- and late-chain, with the latter defined as those coming from isotopes below and including $^{226}$Ra and $^{224}$Ra, respectively. 
}
\smallskip
\resizebox{1.0\linewidth}{!}{
\begin{tabular}{| l | >{\centering\arraybackslash}m{1.0cm} | >{\centering\arraybackslash}m{1.0cm} >{\centering\arraybackslash}m{1.0cm} >{\centering\arraybackslash}m{1.0cm} >{\centering\arraybackslash}m{1.0cm} >{\centering\arraybackslash}m{1.0cm} >{\centering\arraybackslash}m{1.0cm} | >{\centering\arraybackslash}m{1.0cm} | >{\centering\arraybackslash}m{1.0cm} | >{\centering\arraybackslash}m{1.0cm} |}
\hline
\textbf{Background Source} & \textbf{Mass} & \textbf{$^{238}$U$_{e}$} & \textbf{$^{238}$U$_{l}$} & \textbf{$^{232}$Th$_{e}$} & \textbf{$^{232}$Th$_{l}$} & \textbf{$^{60}$Co} & \textbf{$^{40}$K} & \textbf{n/yr} & \textbf{ER} & \textbf{NR} \\ \cline{3-8}
 & \textbf{(kg)} & \multicolumn{6}{|c|}{\textbf{mBq/kg}} & & \textbf{(cts)} & \textbf{(cts)} \\
\hline \hline
\textbf{Detector Components} & & & & & & & & & & \\
PMT systems		& 308   & 31.2 & 5.20 & 2.32 & 2.29 & 1.46 & 18.6 & 248  & 2.82 & 0.027 \\
TPC systems      & 373   & 3.28 & 1.01 & 0.84 & 0.76 & 2.58 & 7.80 & 79.9 & 4.33 & 0.022 \\
Cryostat 	    & 2778  & 2.88 & 0.63 & 0.48 & 0.51 & 0.31 & 2.62 & 323  & 1.27 & 0.018 \\
Outer detector (OD) 	& 22950 & 6.13 & 4.74 & 3.78 & 3.71 & 0.33 & 13.8 & 8061 & 0.62 & 0.001 \\
All else 		& 358   & 3.61 & 1.25 & 0.55 & 0.65 & 1.31 & 2.64 & 39.1 & 0.11 & 0.003 \\
\hline
\multicolumn{9}{|r|}{\textbf{subtotal}} & \textbf{9}  & \textbf{0.07}    \\
\hline \hline
\multicolumn{9}{|l|}{\textbf{Surface Contamination}}            & & \\
\multicolumn{9}{|l|}{Dust (intrinsic activity, 500 ng/cm$^{2}$)}& 0.2  & 0.05 \\
\multicolumn{9}{|l|}{Plate-out (PTFE panels, 50 nBq/cm$^{2}$)}	& -    & 0.05 \\
\multicolumn{9}{|l|}{$^{210}$Bi mobility (0.1 $\si{\micro}$Bq/kg LXe)}		& 40.0 & -	  \\
\multicolumn{9}{|l|}{Ion misreconstruction (50 nBq/cm$^{2}$)}	& -    & 0.16 \\
\multicolumn{9}{|l|}{$^{210}$Pb (in bulk PTFE, 10 mBq/kg PTFE)}		& -	   & 0.12 \\
\hline
\multicolumn{9}{|r|}{\textbf{subtotal}}			& \textbf{40}   & \textbf{0.39} \\
\hline \hline
\multicolumn{9}{|l|}{\textbf{Xenon contaminants}} & & \\
\multicolumn{9}{|l|}{$^{222}$Rn (1.8 $\si{\micro}$Bq/kg)}			 	& 681  & - \\
\multicolumn{9}{|l|}{$^{220}$Rn (0.09 $\si{\micro}$Bq/kg)} 				& 111  & - \\
\multicolumn{9}{|l|}{$^{nat}$Kr (0.015 ppt g/g)}		       		& 24.5 & - \\
\multicolumn{9}{|l|}{$^{nat}$Ar (0.45 ppb g/g)} 	       			& 2.5  & - \\
\hline
\multicolumn{9}{|r|}{\textbf{subtotal}}				& \textbf{819}  & \textbf{0} \\
\hline \hline
\multicolumn{9}{|l|}{\textbf{Laboratory and Cosmogenics}} & & \\
\multicolumn{9}{|l|}{Laboratory rock walls}						& 4.6  & 0.00 \\
\multicolumn{9}{|l|}{Muon induced neutrons}						& -    & 0.06 \\
\multicolumn{9}{|l|}{Cosmogenic activation}						& 0.2  & -    \\
\hline
\multicolumn{9}{|r|}{\textbf{subtotal}}		& \textbf{5}    & \textbf{0.06} \\
\hline \hline
\multicolumn{9}{|l|}{\textbf{Physics}} & & \\
\multicolumn{9}{|l|}{$^{136}$Xe 2$\nu\beta\beta$}               & 67   & - \\
\multicolumn{9}{|l|}{Solar neutrinos: $pp$+$^{7}$Be+$^{13}$N, $^{8}$B$+$$hep$}   							& 191 & 0$^{*}$ \\
\multicolumn{9}{|l|}{Diffuse supernova neutrinos (DSN)}	          	& -    & 0.05 \\
\multicolumn{9}{|l|}{Atmospheric neutrinos (Atm)}		            	& -    & 0.46 \\
\hline
\multicolumn{9}{|r|}{\textbf{subtotal}}							& \textbf{258}  & \textbf{0.51} \\
\hline \hline
\multicolumn{9}{|l|}{Total}                                     & 1131 & 1.03 \\
\multicolumn{9}{|l|}{Total (with 99.5$\%$ ER discrimination, 50$\%$ NR efficiency)} & 5.66 & 0.52 \\
\hline
\multicolumn{9}{|l|}{\textbf{Sum of ER and NR in LZ for 1000 days, 5.6 tonne FV, with all analysis cuts}} & \multicolumn{2}{|c|}{\textbf{6.18}} \\
\hline
\multicolumn{11}{l}{{\footnotesize* Below the 6~keV NR threshold used here.}}
\end{tabular}
}

\label{table:backgrounds}
\end{table*}

\subsection{Trace radioactivity in detector components}

The most prevalent isotopes in naturally-occurring radioactive materials (NORMs) are the gamma-emitting isotopes $^{40}$K, $^{137}$Cs and $^{60}$Co, as well as $^{238}$U, $^{235}$U, $^{232}$Th and their progeny. The TDR~\cite{Mount:2017qzi} describes the facilities utilized to measure the radioactivity of detector materials, and LZ is undertaking a campaign involving nearly 2000 radio-assays of the materials that form the composite assemblies, components or sub-components listed in Table~\ref{table:backgrounds}. As a result of this comprehensive program and the power of self-shielding afforded by LXe, trace radioactivity in detector materials is not expected to be a leading cause of background to the experiment.

\subsection{Surface contaminants}

Radioactivity on detector surfaces arises from the accumulation of $^{222}$Rn-daughters plated-out during the manufacture and assembly of components, as well as generic dust contamination containing NORMs that release gamma rays and induce neutron emission. Plate-out can generate NR backgrounds through two mechanisms: ($\alpha,n$) processes that release neutrons into the xenon; and ions from the $^{210}$Pb sub-chain originating at the edge of the TPC being misreconstructed as NRs within the fiducial volume. The impact of the latter depends critically on the performance of position reconstruction and drives the 4~cm radial fiducial volume cut (see Sec.~\ref{sec:analysiscuts}). LZ has instituted a target for plate-out of $^{210}$Pb and $^{210}$Po of less than 0.5~mBq/m$^{2}$ on the TPC walls and below 10~mBq/m$^{2}$ everywhere else. LZ has also instituted a requirement limiting generic dust contamination to less than 500~ng/cm$^{2}$ on all wetted surfaces in the detector and xenon circulation system.
A rigorous program of cleanliness management is implemented to ensure that the accumulated surface and dust contamination do not exceed these limits. All detector components that contact xenon must be cleaned and assembled according to validated cleanliness protocols and witness plates will accompany the production and assembly of all detector components. Detector integration will take place in a reduced-radon cleanroom built at the Surface Assembly Laboratory at SURF.

Several large volume liquid scintillator experiments reported observing mobility of radon-daughters plated onto surfaces, in particular the beta emitter $^{210}$Bi~\cite{Gando:2014wjd,Takemoto:2015gta,Bellini:2013lnn}. Studies in LUX are used to place a limit on this mobility in LXe, resulting in the projection shown in Table~\ref{table:backgrounds}. 

\subsection{Dispersed xenon contaminants}

Radioisotopes dispersed throughout the LXe produce background that cannot be mitigated through self-shielding. Radon emanation from materials and dust results in the largest contribution to the total background in LZ. This is primarily due to `naked beta' emission---a beta emitted without any accompanying gamma rays---from $^{214}$Pb($^{212}$Pb) in the $^{222}$Rn($^{220}$Rn) sub-chain.
To simulate the radon contribution to Table~\ref{table:backgrounds}, the default branching ratios in GEANT4 are modified: that from $^{214}$Pb($^{212}$Pb) to the ground state of $^{214}$Bi($^{212}$Bi) is taken to be 9.2\%(13.3\%)~\cite{pb214nucleide:2010,pb212nucleide:2011}.
Direct measurements of $^{222}$Rn emanation from xenon-wetted materials are performed~\cite{Miller:2017tpl}. For components that do not yet exist or are still to be measured, projections are made based on measurements of similar materials that exist in the literature. Most measurements are made at room temperature, and the expected emanation can depend strongly on temperature depending on the source material. For these estimates a conservative approach is adopted, only taking credit for a reduction at LXe temperatures if there is direct knowledge that such a reduction will occur. The LZ gas handling apparatus~\cite{Mount:2017qzi} includes a radon reduction system that can take a small stream of gas from problem areas, such as the cable conduits, and perform on-line radon purification~\cite{Pushkin:2018wdl}. The current best estimate for emanation from LZ components results in a $^{222}$Rn specific activity of 1.53~$\si{\micro}$Bq/kg of LXe. 

Radon emanation from dust is estimated separately. For the radioactivity levels typical of dust at SURF and under the conservative assumption, compared to preliminary measurements, that 25\% of $^{222}$Rn is released into the LXe, the dust requirement of $<$500~ng/cm$^{2}$ generates a $^{222}$Rn specific activity of 0.28~$\si{\micro}$Bq/kg of LXe. Combined with the emanation from detector components, a total of 1.8~$\si{\micro}$Bq/kg of $^{222}$Rn is projected. A concentration of 0.09~$\si{\micro}$Bq/kg of $^{220}$Rn ($\times$0.05 the specific activity of $^{222}$Rn, based on the ratio seen in LUX~\cite{akerib:2014rda}) is also included in the background estimates.

Natural xenon includes trace levels of \IKreF and $^{39}$Ar, both of which disperse throughout the liquid and are beta emitters that lead to ER events in the ROI. LZ has instituted a significant xenon purification campaign using chromatography to remove krypton from xenon in order to control \IKreF. In an R\&D phase, the chromatography system reduced the $^{\mathrm{nat}}$Kr/Xe concentration to 0.075~ppt~g/g~\cite{Mount:2017qzi} and a further improvement to 0.015 ppt~g/g is expected in the production system. Argon levels are also reduced during this purification step, with an expected concentration of $^{\mathrm{nat}}$Ar/Xe below 0.45~ppb~g/g. 

\subsection{Laboratory and cosmogenic backgrounds}

Neutrons produced from muon-induced electromagnetic and hadronic cascades can generate background events~\cite{reichhart:2013xkd,gray:2010nc}. The number of muon-induced NR background events has been estimated using simulations of muon transport through rock around the laboratory and detector geometry, including secondary particle production, transport and detection. Backgrounds from outside the water tank are dominated by the cavern walls. The gamma flux has been measured at the 4850-foot level of SURF (4300~m~w.e.) at various locations in the Davis Campus~\cite{mei:2009py,thomas:2014}. Neutrons from the laboratory walls are attenuated efficiently by water and scintillator surrounding the LZ cryostat; with a minimum thickness of hydrogenous shielding of 70 cm, the neutron flux is reduced by more than 6 orders of magnitude~\cite{Tomasello:2010zz} resulting in a negligible contribution to backgrounds in LZ.

Cosmogenic activation of xenon can lead to contamination by $^{127}$Xe (T$_{1/2} = 36.4$~d). LUX measurements~\cite{akerib:2014rda} show an equilibrium decay rate of $(2.7 \pm 0.5)$~mBq/kg of $^{127}$Xe after xenon was exposed to cosmic rays on the Earth's surface (see also~\cite{Baudis:2015kqa}). That level of activity leads to the projected number of events shown in Table~\ref{table:backgrounds} following an assumed 8-month cooling down period underground prior to data-taking.  The largest contribution to activation in the detector materials comes from production of $^{46}$Sc (T$_{1/2} = 83.8$~d) in the 2.5~tonnes of titanium being used in LZ. Using GEANT4 and ACTIVIA~\cite{back:2007kk,Zhang:2016rlz} simulations, the decay rate of $^{46}$Sc is estimated to be 4.8~mBq/kg of titanium after 6~months activation at sea level and surface assembly of the TPC within the cryostat at SURF, followed by the same 8~month cooling down period underground assumed for $^{127}$Xe.

\subsection{Physics backgrounds} 
Three sources of background are identified that carry interesting physics in their own right: neutrino-electron scattering (ER), 2$\nu\beta\beta$ $^{136}$Xe decay (ER), and neutrino-nucleus scattering (NR).  All three of these backgrounds generate single-scatter events uniformly in the detector with no corresponding veto signal.

The solar neutrino ER background is dominated by $pp$ neutrinos, with smaller contributions from  the $^7$Be, and CNO chains, and LZ uses the flux and spectra from~\cite{bahcall:2004mz} and up to date oscillation parameters from~\cite{Olive:2016xmw} to calculate the solar neutrino rates. Below 30~keV a scaling factor is applied to the free electron scattering rate based on the relativistic random phase approximation calculation in~\cite{Chen:2017eab} that accounts for the effect of atomic binding. The rate of 2$\nu\beta\beta$ decay of $^{136}$Xe in LZ is based on measurements in EXO-200 and KamLAND-Zen~\cite{ackerman:2011gz,kamlandzen:2012aa}.

Nuclear recoils are produced by $^8$B and \emph{hep} solar neutrinos, diffuse supernova neutrinos and atmospheric neutrinos through coherent elastic neutrino-nucleus scattering, a standard model process that was recently observed for the first time~\cite{Akimov:2017ade}. $^8$B- and \emph{hep}-induced events populate the very low recoil energy region, and their impact depends critically on the NR efficiency shown in Fig.~\ref{fig:nr_er_efficiency}. Because these neutrino fluxes do not constitute a significant background in searches for WIMPs of mass $> 20$ GeV, they are not included in Table~\ref{table:backgrounds}; however, they are included in the WIMP sensitivity calculations using the full PLR treatment described in section~\ref{sec:WIMPsens}. Atmospheric and diffuse supernova neutrinos produce NRs at higher energies and constitute the largest contribution to the total NR background in LZ. When modelling atmospheric neutrinos the Gran~Sasso flux from~\cite{battistoni:2005pd} is used with a correction to account for the difference in geomagnetic cutoff rigidity~\cite{smart20052012} at SURF.

\subsection{Non-standard backgrounds}

\begin{figure*}[t!]
\begin{tabular}{ccc}
\includegraphics[trim={0.0cm 0.0cm 0.0cm 0.0cm},clip,width=0.45\linewidth]{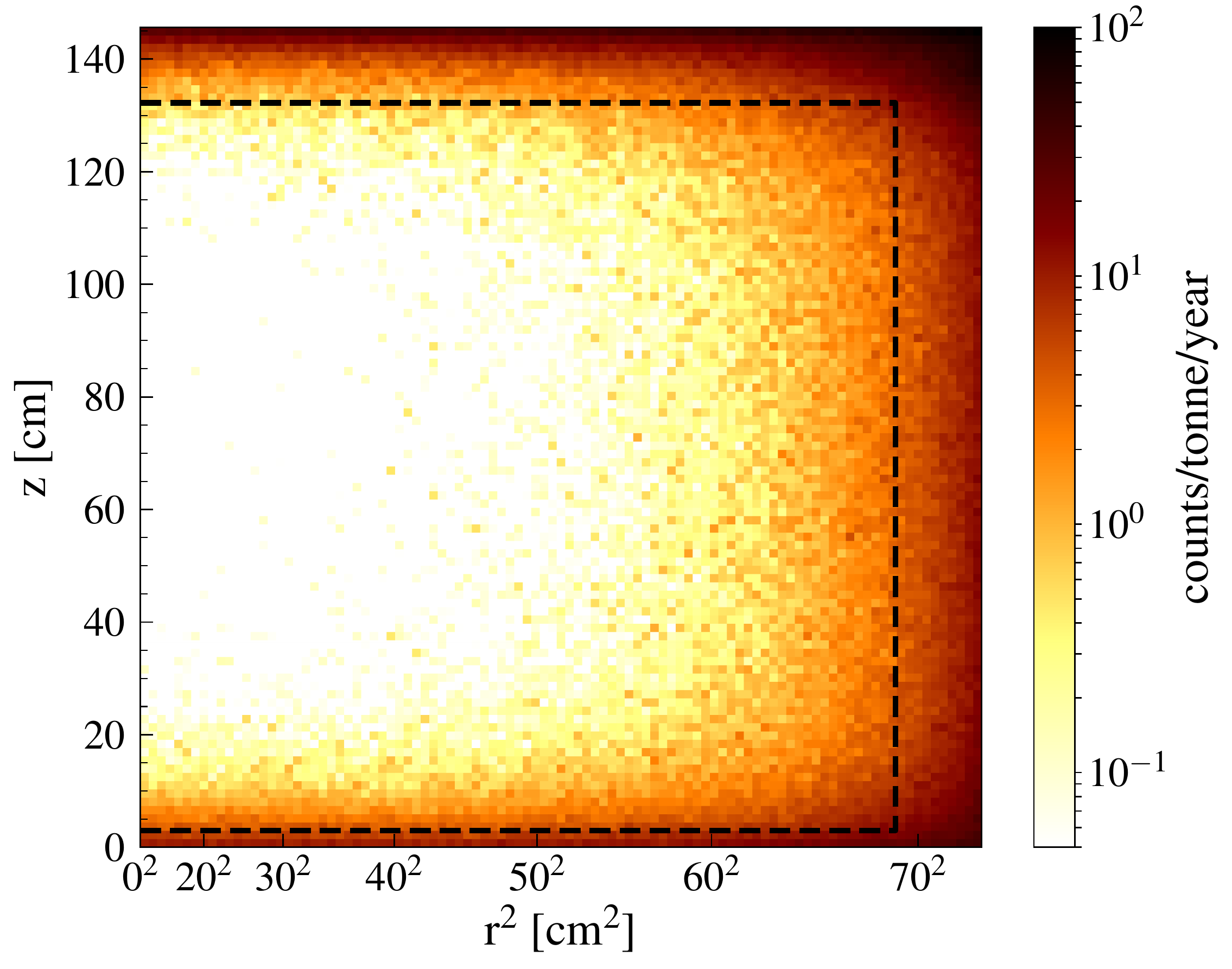} & &
\includegraphics[trim={0.0cm 0.0cm 0.0cm 0.0cm},clip,width=0.45\linewidth]{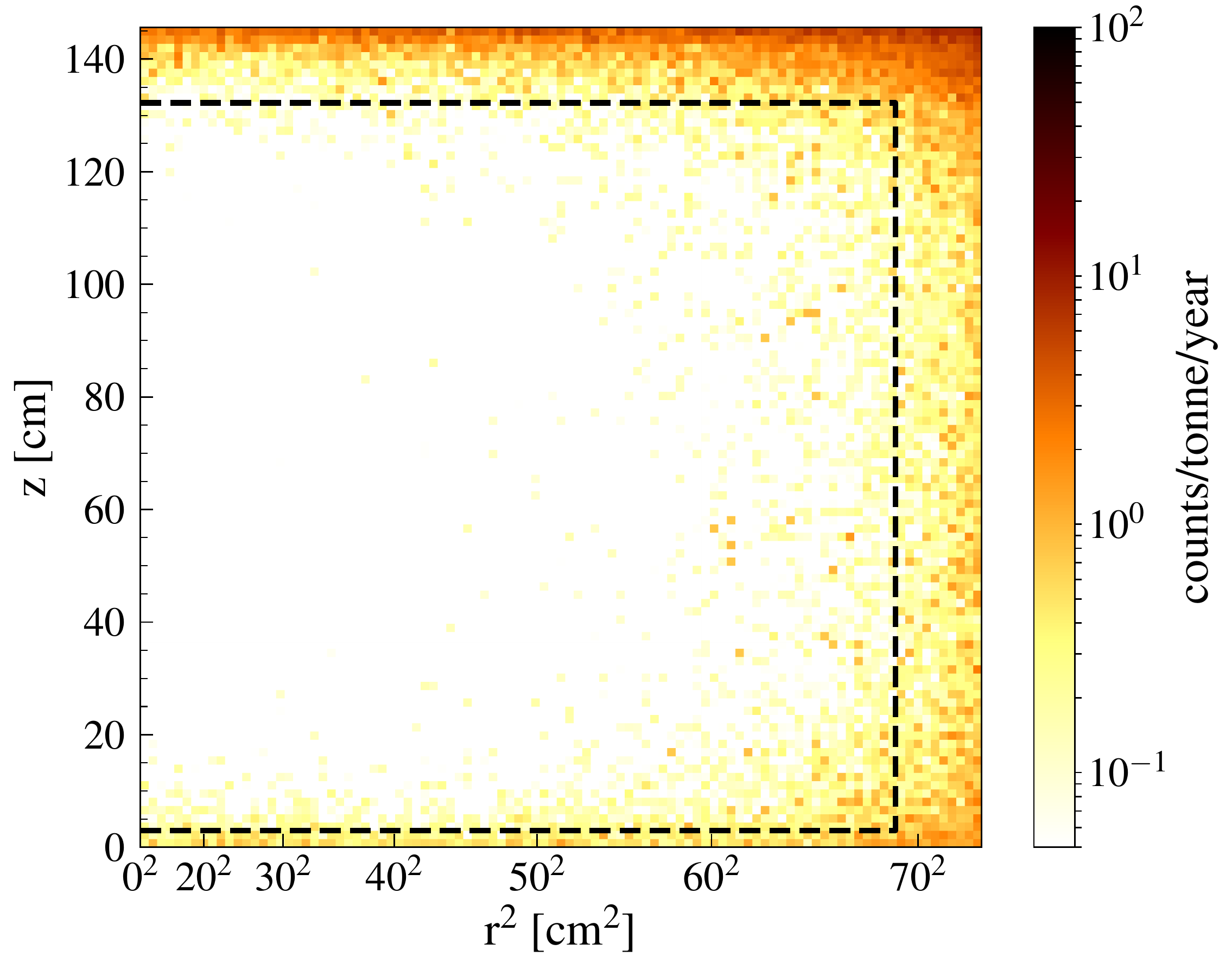} \\
\end{tabular}
\caption{Single scatter event distributions for all significant NR backgrounds in the region of interest relevant to a 40~GeV/c$^{2}$ WIMP (approximately 6--30~keV) with no vetoing (left) and after application of both xenon skin and OD vetoes (right). To not obscure the spatial dependence within the fiducial volume contributions from the uniform low-energy $^{8}$B and \emph{hep} events and the sharply falling radial wall events are omitted. The integrated counts for all NR backgrounds in the 5.6 tonne fiducial volume (dashed line) are reduced from 10.4~cts/1000~days with no vetoing to 1.03~cts/1000~days after application of the vetoes.
}
\label{fig:z2_vs_r_temp_plots_nr}
\end{figure*}

A number of rare but potentially dangerous non-standard event topologies are considered. While not currently included as components in the background model used for sensitivity projections, studies have been performed to ensure they are sub-dominant to the existing ER and NR backgrounds.

Multiple scattering of gamma rays where one vertex occurs in a region of the detector that is optically coupled to the PMT arrays but that has no charge collection can cause an NR-like background. These so-called `gamma-X' events have a lower S2/S1 ratio than is typical for ER events and can leak into the NR band. Simulations indicate that with the current fiducial volume (2~cm from cathode) less than 0.1 of these events are expected. 

Accidental coincidences between multiple PMT dark counts lead to a rate of fake S1-only signals; these may combine with S2-only events to fake plausible S1-S2 pairs, some of which can overlap with the NR band. Considering the cold PMT dark count measurements reported in~\cite{Aprile:2017aty}, enforcing a 3-fold PMT coincidence level for a valid S1 signal, and predicting 1~mHz for the S2-only rate (twice that seen in LUX~\cite{Akerib:2015rjg}) then less than 0.2~events are projected in a 1000~day run.

S1-like signals from Cherenkov light generated in the PMT quartz windows (e.g.~from energetic betas or Compton electrons from $^{40}$K decays internal to the PMTs~\cite{araujo:2011as}) were also considered. Such signals can combine with S2-only events to create fake S1-S2 pairs that populate the WIMP search region of interest as low-energy NR-like events. Fortunately, the majority of these Cherenkov signals can be readily identified based on their timing and PMT hit patterns, typically possessing a spread in arrival times of less than 10~ns with the majority of the light detected in the source PMT. These characteristics and the above S2-only rate lead to a projection of 0.2~events in a 1000~day run.

\subsection{Spatial distribution of NR backgrounds and effect of the vetoes}

The spatial distribution of single scatter NR events from all significant background sources is shown in Fig.~\ref{fig:z2_vs_r_temp_plots_nr} before (left) and after (right) application of the veto detectors. Neither the low-energy $^{8}$B and \emph{hep} events nor the sharply falling radial wall events are included in Fig.~\ref{fig:z2_vs_r_temp_plots_nr}. Without the veto system, the rate of NR events inside the fiducial volume increases by a factor of around 10 from 1.03~cts/1000~days to 10.4~cts/1000~days, severely impacting the sensitivity and discovery potential of LZ. A reduction in fiducial mass to approximately 3.2~tonnes would be necessary to reduce the NR rate  to that achievable with the veto system and the full 5.6~tonne fiducial mass.

\section{WIMP Sensitivity}
\label{sec:WIMPsens}
The LZ projected sensitivity to SI and SD WIMP-nucleon scattering is calculated for an exposure of 1000~live days and a fiducial mass of 5.6 tonnes. The sensitivity is defined as the median 90\% confidence level (CL) upper limit on the relevant WIMP-nucleon cross section that would be obtained in repeated experiments given the background-only hypothesis. It is evaluated using the Profile Likelihood Ratio (PLR) method~\cite{rolke:2004mj} using an unbinned and extended likelihood~\cite{Olcina:2019} that provides near-optimal exploitation of the differences between signal and background, based on the position-corrected signals S1$_c$ and S2$_c$. For these projections no position information is included in the list of PLR observables and instead the simple cylindrical fiducial volume cut described in Sec.~\ref{sec:Sims} is applied, containing 5.6 tonnes of LXe. A scan over cross section is performed for each WIMP mass, and the 90\% confidence interval is obtained by performing a frequentist hypothesis test inversion using the RooStats package~\cite{moneta:2010pm}. For the limit projections shown here, a one-sided PLR test statistic for upper limits is used, cf.~equation (14) in~\cite{Cowan:2010js}, where to protect against the exclusion of cross sections for which LZ is not sensitive the result will be power constrained between $-1\sigma$ and the median sensitivity following~\cite{Cowan:2011an} (by construction the median expected sensitivity is unchanged); for evaluating discovery potential a test statistic for rejecting the null hypothesis is used, following equation (12) in~\cite{Cowan:2010js}. For limit projections Monte Carlo calculations are used to evaluate the test statistic but for discovery projections the asymptotic formulae derived in~\cite{Cowan:2010js} are used. 

An 11-component background model is built for the PLR based on the estimates described in Sec.~\ref{sec:BGs} and shown in Table~\ref{table:plr_inputs}. Contributions from detector components, surface contamination and environmental backgrounds are summed together into a single \emph{Det.~+~Sur.~+~Env.} component. Also shown in Table~\ref{table:plr_inputs} are systematic uncertainties on the normalization of each background. The uncertainties on the \emph{Det.~+~Sur.~+~Env.} component are estimated from the counting and simulation results, those on the neutrino components are primarily flux uncertainties, those on the radon contribution come from uncertainty in the branching ratio of $^{214}$Pb and $^{212}$Pb to their respective ground states, and those on $^{85}$Kr and $^{136}$Xe from uncertainty on the spectral shapes at low energies. These systematics are treated as nuisance terms with Gaussian priors in the PLR calculation, but they do not have a significant effect on the sensitivity because of the low number of background counts expected in LZ. No other nuisance terms are included in the sensitivity calculation presented here. 

\begin{table}[t!]
\centering
\caption{Eleven background types considered in the PLR analysis, along with the integrated counts in the LZ 1000~day WIMP search exposure and the systematic uncertainties on their normalizations, included as nuisance parameters in the PLR. Counts are for the WIMP search ROI (S1 with $\geq 3\mathrm{-fold}$ coincidence, S1$_{c}<80$~phd and uncorrected S2 $> 415$~phd): approximately 1.5--15~keV for ERs and 4--60~keV for NRs; and after application of the single scatter, skin and OD veto, and 5.6~tonne fiducial volume cuts.}
\begin{tabular}{ | r | c | c | }
    \hline
    Background & N & $\sigma/N$ \\
    \hline
    \hline
    $^{222}$Rn (ER) & 1915 & 10\% \\
    $pp$+$^{7}$Be+$^{14}$N $\nu$ (ER) & 615 & 2\% \\
    $^{220}$Rn (ER) & 316 & 10\% \\
    $^{136}$Xe 2$\nu\beta\beta$ (ER) & 495 & 50\% \\ 
    Det. $+$ Sur. $+$ Env. (ER)  & 171 & 20\% \\
    $^{85}$Kr (ER)  & 83 & 20\% \\
    $^{8}$B solar $\nu$ (NR) & 36 & 4\% \\
    Det. $+$ Sur. $+$ Env. (NR) & 0.81 & 20\% \\
    Atmospheric $\nu$ (NR)  & 0.65 & 25\% \\
    \emph{hep} $\nu$ (NR)  & 0.9 & 15\% \\
    DSN $\nu$ (NR)  & 0.15 & 50\% \\
    \hline
\end{tabular}
\label{table:plr_inputs}
\end{table}

\begin{figure}[b!]
\centering
\includegraphics[trim={0.6cm 0.5cm 0.3cm 0.2cm},width=0.88\linewidth]{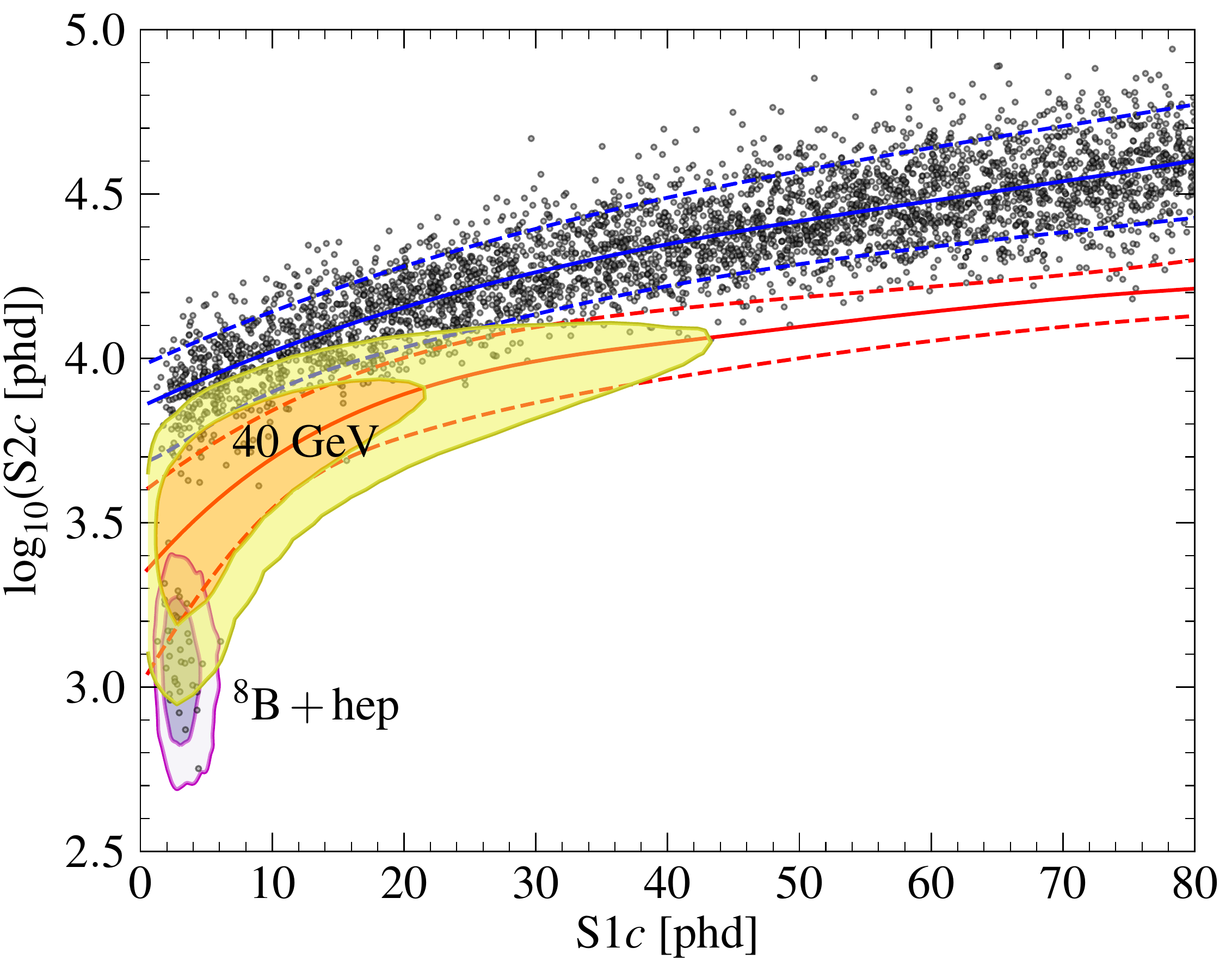}
\caption{LZ simulated data set for a background-only 1000~live day run and a 5.6~tonne fiducial mass. ER and NR bands are indicated in blue and red, respectively (solid: mean; dashed: 10\% and 90\%). The 1$\sigma$ and 2$\sigma$ contours for the low-energy $^{8}$B and \emph{hep} NR backgrounds, and a 40~$\mathrm{GeV}/c^{2}$ WIMP are shown as shaded regions.
}
\label{fig:example_sig_bkg_pdfs}
\end{figure}

The signal spectrum for WIMP recoils is calculated using the standard halo model 
following the formalism of~\cite{mccabe:2010zh}, with $\upsilon_{0} = 220$~km/s; $\upsilon_{esc} = 544$~km/s; $\upsilon_{e} = 230$~km/s and $\rho_{0} = 0.3$~GeV/$c^{2}$. For SI scattering the Helm form factor~\cite{Helm:1956zz} is used as in~\cite{lewin:1995rx}, while for SD scattering structure functions are taken from~\cite{klos:2013rwa}. For these projections no uncertainty is assumed in the signal model. 
Signal and background PDFs in S1$_c$ and S2$_c$ are created using NEST and the parameterization of detector response described in Sec.~\ref{sec:Sims} and shown in Table~\ref{table:parameters}. 
The power of the PLR technique arises from an optimal weighting of the background-free and background-rich regions, and for all WIMP masses considered background rejection exceeds 99.5\% for a signal acceptance of 50\%. Figure~\ref{fig:example_sig_bkg_pdfs} demonstrates the separation in (S1$_{c}$,S2$_{c}$) of a 40~$\mathrm{GeV}/c^{2}$ WIMP signal from the LZ backgrounds expected in a 1000~day run.

\subsection{Spin-independent scattering}
\label{sec:spin-independent-scattering}

\begin{figure}[t!]
\centering
\includegraphics[trim={0.2cm 0.3cm 1.3cm 0.4cm},clip,width=0.98\linewidth]{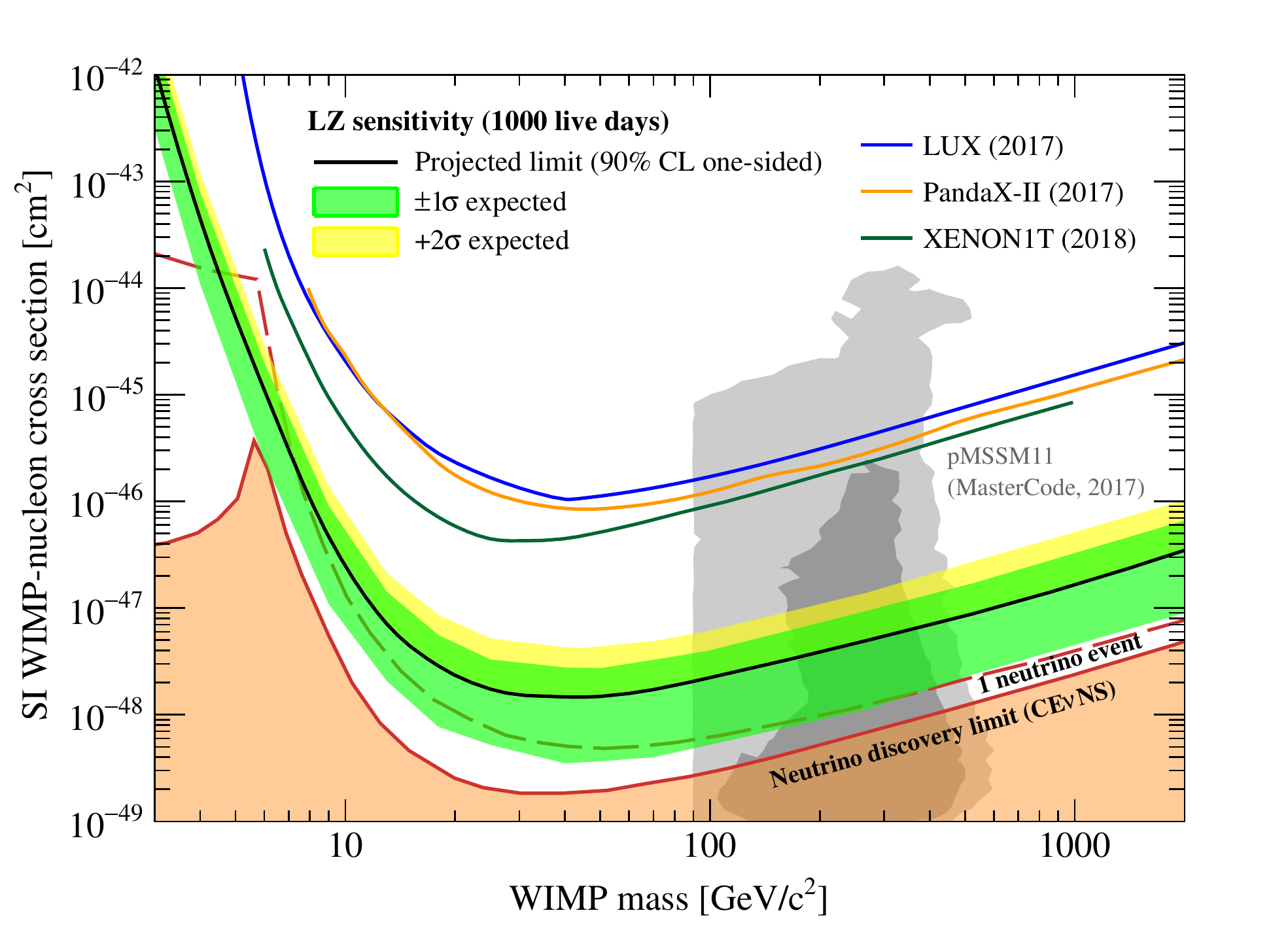}
\caption{LZ projected sensitivity to SI WIMP-nucleon elastic scattering for 1000~live days and a 5.6~tonne fiducial mass. The best sensitivity of $1.4 \times 10^{-48}$~cm$^{2}$ is achieved at a WIMP mass of 40~$\mathrm{GeV}/c^{2}$. The $-2\sigma$ expected region is omitted based on the expectation that the limit will be power constrained~\cite{Cowan:2011an}. Results from other LXe experiments are also shown~\cite{Akerib:2016vxi,Cui:2017nnn,Aprile:2018dbl}. The lower shaded region and dashed line indicate the emergence of backgrounds from coherent scattering of neutrinos~\cite{billard:2013qya,ruppin:2014bra} and the gray contoured regions show the favored regions from recent pMSSM11 model scans~\cite{Bagnaschi:2017tru}.} 
\label{fig:wimp_si_sensitivity_vs_mass}
\end{figure}

The LZ projected sensitivity to SI WIMP-nucleon scattering is shown in Fig.~\ref{fig:wimp_si_sensitivity_vs_mass}. A minimum sensitivity of $1.4 \times 10^{-48}$~cm$^{2}$ is expected for 40~$\mathrm{GeV}/c^{2}$ WIMPs, more than an order of magnitude below the limits set by recent LXe experiments. With this sensitivity LZ will probe a significant fraction of the parameter space remaining above the irreducible background from coherent scattering of neutrinos from astrophysical sources, intersecting several favored model regions on its way. 

The higher light collection efficiency compared to the baseline presented in the TDR~\cite{Mount:2017qzi} (from 7.5\% to 11.9\%) leads to an improvement at all WIMP masses. 
The lower energy threshold leads to a significant expected rate of coherent neutrino-nucleus scattering from \IBe and \emph{hep} neutrinos, with $36$ and $0.9$ counts expected in the full exposure, respectively.
These events are not a background at most WIMP masses but are interesting in their own right and would constitute the first observation of coherent nuclear scattering from astrophysical neutrinos.

The observed rate of events from \IBe and \emph{hep} neutrinos as well as sensitivity to low mass WIMPs will depend strongly on the low energy nuclear recoil efficiency (see Fig.~\ref{fig:nr_er_efficiency}). Recent results from LUX and XENON1T appropriately assume a cutoff in signal below 1.1~keV to obtain conservative upper limits~\cite{Akerib:2016vxi,Aprile:2017iyp}, even though such a cutoff is not physically motivated. The results shown here are projections only, and an extrapolation down to 0.1~keV following Lindhard theory is used. Use of a hard cutoff at 1.1~keV would degrade sensitivity to a 4~GeV/$c^{2}$ mass WIMP by a factor of two, with no significant effect on sensitivity to WIMP masses above 6~GeV/$c^{2}$. The expected rate of \IBe background events would also decrease by about 20\%. Ultimately, the planned suite of low energy nuclear recoil calibrations will be needed to fully characterize the sensitivity of LZ to low mass WIMP and \IBe neutrino signals.

\begin{figure}[b!]
\centering
\includegraphics[trim={0.2cm 0.2cm 1.3cm 0.3cm},clip,width=0.98\linewidth]{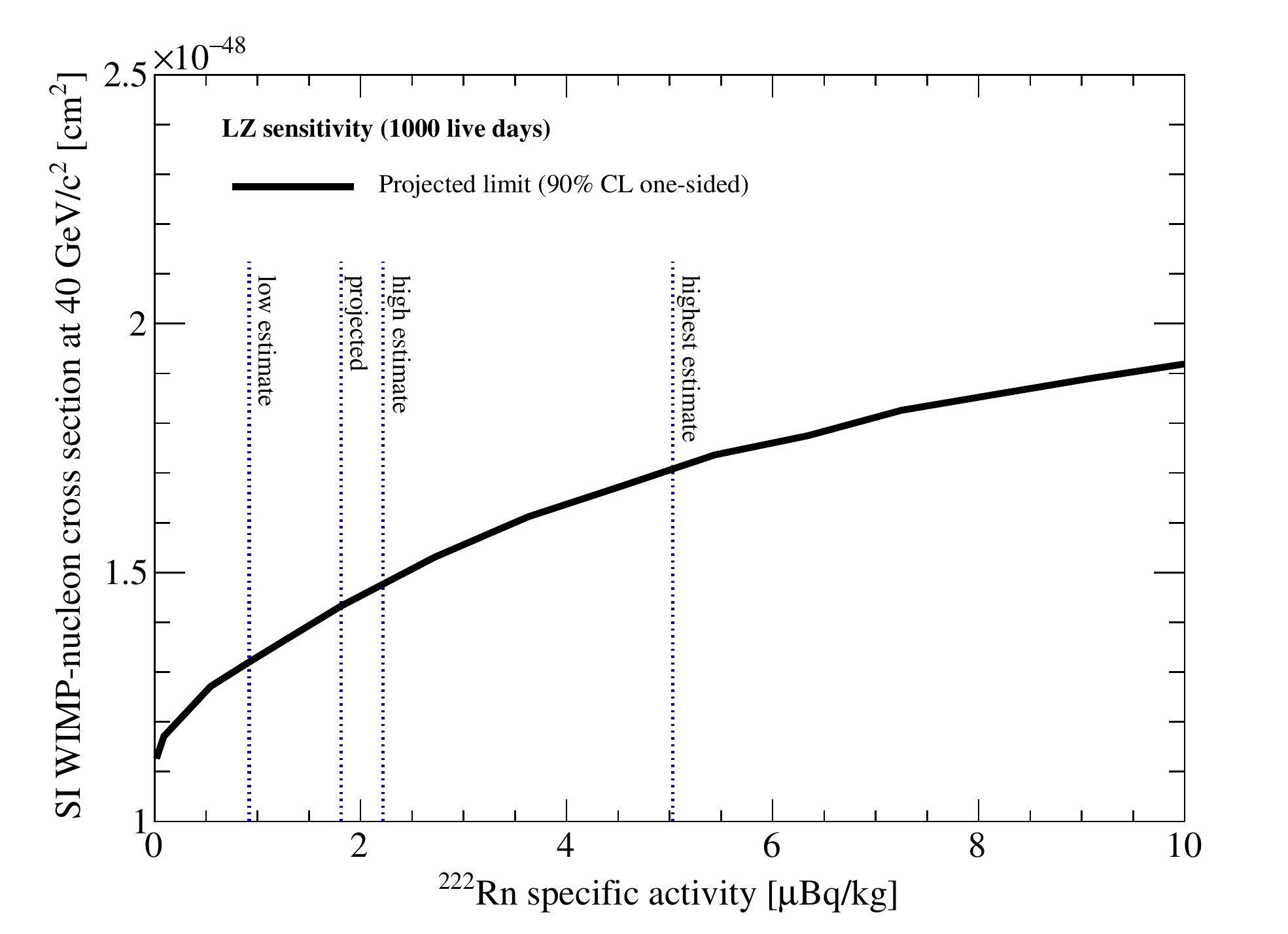}
\caption{LZ projected SI sensitivity for a 40~$\mathrm{GeV}/c^{2}$ WIMP as a function of overall Rn level, for a 5.6~tonne fiducial volume and a 1000 live day run. Included in the variation is an implicit $^{220}$Rn contribution at $1/20$th the specific activity of $^{222}$Rn. The dashed vertical lines indicate the various Rn scenarios. 
}
\label{fig:sensitivity_vs_rn_level}
\end{figure}

Since radon is projected to be the largest source of events, a number of scenarios are considered based on current assessments for radon rates in LZ: the nominal \textit{projected} scenario (1.8~$\si{\micro}$Bq/kg of $^{222}$Rn with an implicit $^{220}$Rn contribution at $1/20$th the specific activity of $^{222}$Rn); a \textit{high estimate} (2.2~$\si{\micro}$Bq/kg) and \textit{low estimate} (0.9~$\si{\micro}$Bq/kg) that correspond to all Rn-screening measurements being aligned at their $+ 1\sigma$ and $- 1\sigma$ expectations, respectively; and a \textit{highest estimate} scenario (5.0~$\si{\micro}$Bq/kg) that in addition to $+ 1\sigma$ expectations also assumes no reduction in emanation rate at LZ operating temperatures. Figure~\ref{fig:sensitivity_vs_rn_level} shows how the SI sensitivity to a 40~$\mathrm{GeV}/c^{2}$ WIMP varies as a function of overall radon concentration in the 5.6~tonne fiducial volume. Even for the \textit{highest estimate} scenario the median sensitivity is better than~$3 \times 10^{-48}$~cm$^{2}$. Scans of sensitivity as a function of other background components and as a function of several detector parameters can be found in~\cite{Mount:2017qzi}. 

\subsection{Discovery potential}

\begin{figure}[t!]
\centering
\includegraphics[trim={0.0cm 0.1cm 1.5cm 0.6cm},clip,width=0.98\linewidth]{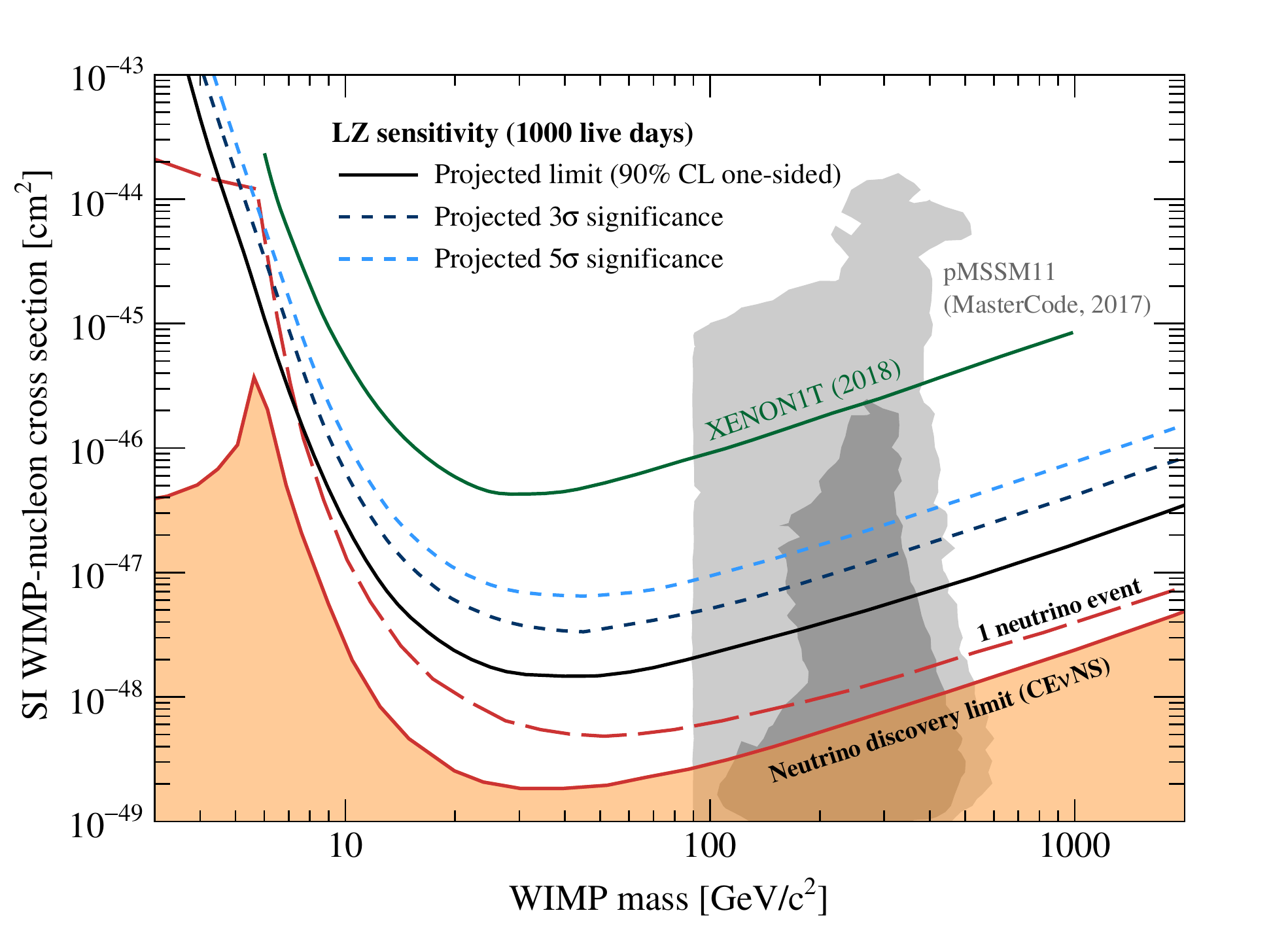}
\caption{LZ discovery potential for SI WIMP-nucleon scattering. The best 3(5)$\sigma$ significance is achieved at $3.4(6.5) \times 10^{-48}$~cm$^{2}$ for 40~GeV/$c^{2}$ WIMPs. The current best limit (XENON1T~\cite{Aprile:2018dbl}) is shown for comparison.  
}
\label{fig:projected_SI_discovery}
\end{figure}

LZ discovery potential for SI WIMP-nucleon scattering is shown in Fig.~\ref{fig:projected_SI_discovery}, where the ability to exclude the null result at 3$\sigma$ and 5$\sigma$ significance is shown as a function of WIMP mass and is compared to existing and future LXe 90\% CL sensitivities. At 40~GeV/$c^{2}$ the median 3(5)$\sigma$ significance will occur at $3.4(6.5) \times 10^{-48}$~cm$^{2}$. For all WIMP masses the projected 5$\sigma$ significance is below the 90\%~CL limits from recent experiments.

\subsection{Spin-dependent scattering}

\begin{figure*}[t!]
\begin{tabular}{ccc}
\includegraphics[trim={0.1cm 0.0cm 1.3cm 0.4cm},clip,width=0.49\linewidth]{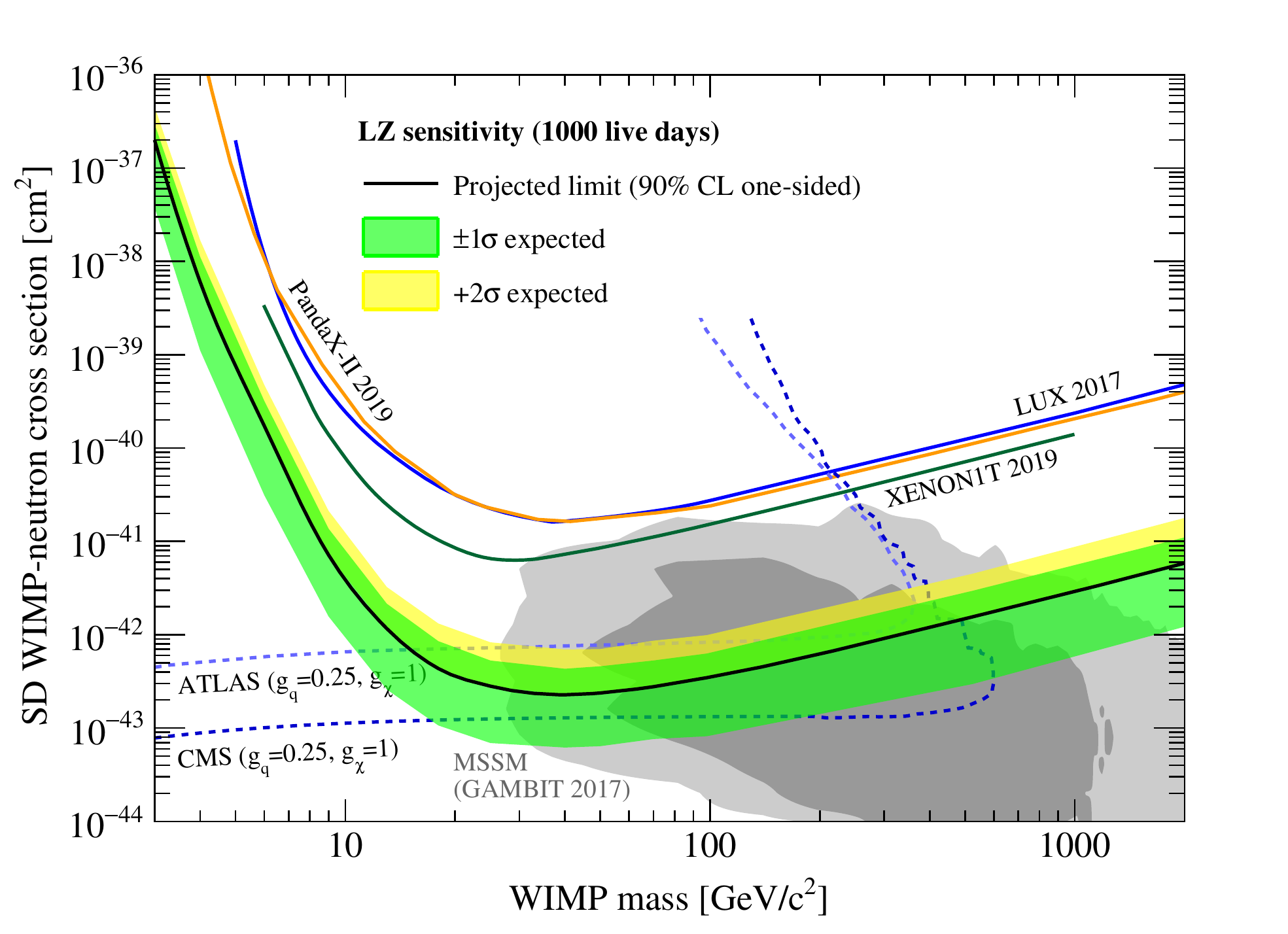} & &
\includegraphics[trim={0.1cm 0.0cm 1.3cm 0.4cm},clip,width=0.49\linewidth]{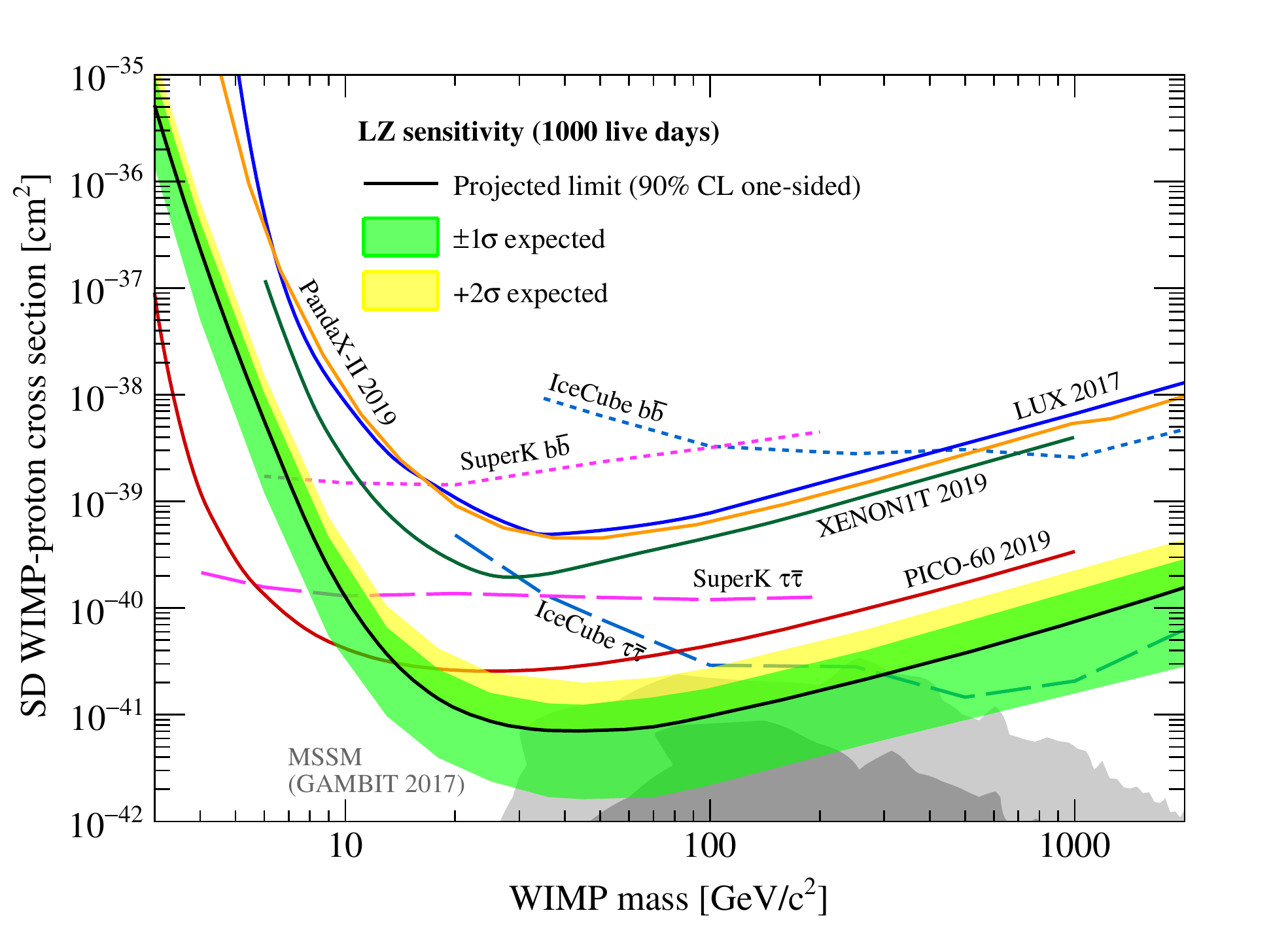} \\
\end{tabular}
\caption{LZ projected sensitivity to SD WIMP-neutron (left) and WIMP-proton (right) scattering for a 1000~live day run with a 5.6 tonne fiducial mass. For SD WIMP-neutron(-proton) scattering a minimum sensitivity of $2.3 \times 10^{-43}$~cm$^{2}$ ($7.1 \times 10^{-42}$~cm$^{2}$) is expected at 40~$\mathrm{GeV}/c^{2}$. Recent limits from direct detection experiments are shown as solid lines~\cite{Akerib:2017kat,Xia:2018qgs,Aprile:2019dbj,Amole:2019fdf}. Dashed lines indicate the model dependent collider constraints from the LHC (for WIMP-neutron)~\cite{Aaboud:2017dor, Sirunyan:2017hci} and the leading indirect limits from neutrino telescopes (for WIMP-proton)~\cite{Choi:2015ara, Aartsen:2016zhm}. The gray contoured regions show the favored regions from recent MSSM7 model predictions~\cite{Athron:2017yua}.
}
\label{fig:wimp_sd-n_sensitivity_vs_mass}
\end{figure*}

The sensitivity of LZ to SD WIMP-neutron and WIMP-proton scattering is shown in Fig.~\ref{fig:wimp_sd-n_sensitivity_vs_mass}. Naturally occurring xenon has an abundance of around 50\% in isotopes with odd neutron number (26.4\%~$^{129}$Xe and 21.2\%~$^{131}$Xe by mass). For SD WIMP-neutron(-proton) scattering a minimum sensitivity of $2.3 \times 10^{-43}$~cm$^{2}$ ($7.1 \times 10^{-42}$~cm$^{2}$) is expected at 40~$\mathrm{GeV}/c^{2}$. LZ will explore a significant fraction of the favored MSSM7 model region~\cite{Athron:2017yua} for SD WIMP-neutron scattering.

\section{Conclusions}

The physics run of LZ, starting in 2020, will probe a significant fraction of the remaining parameter space for the direct detection of WIMPs.

The LZ detector has been designed to maximize target mass and exposure, while achieving ultra-low radioactivity and active monitoring of residual backgrounds. The outer detector and active xenon skin veto systems are critical to this: providing both the rejection of neutrons and gamma rays from internal sources and the characterization of the environmental backgrounds in the vicinity of the core TPC, to give a powerful \emph{in situ} constraint on the rates of processes that might produce backgrounds to WIMP signals.

The sensitivity of LZ has been evaluated with a detector response built on the properties of the materials procured for use in LZ and a background model based on the results of a comprehensive materials screening campaign. 

For a 1000 day exposure utilizing a 5.6~tonne fiducial mass, LZ is projected to exclude, at 90\% CL, SI WIMP-nucleon cross sections of $1.4 \times 10^{-48}$~cm$^{2}$ and above for a 40~$\mathrm{GeV}/c^{2}$ WIMP. This represents over an order of magnitude improvement with respect to the final sensitivities of recent LXe dark matter experiments; LZ will have $5\sigma$ discovery potential for cross sections below their expected 90\% exclusion limits giving access to an entirely unexplored class of theoretical models and predictions~\cite{cahillrowley:2014boa}. For SD WIMP-neutron (-proton) scattering, a best sensitivity of $2.3 \times 10^{-43}$~cm$^{2}$ ($7.1 \times 10^{-42}$~cm$^{2}$) for a 40~$\mathrm{GeV}/c^{2}$ WIMP is expected.

Underground installation of LZ is now well underway and the experiment is on track for commissioning at SURF in 2020.

\section{Acknowledgements}

We acknowledge the important contribution of our deceased colleague Prof. James 
White of Texas A\&M University, whose vision was fundamental to the conceptual 
design and experimental strategy of LZ. 

The research supporting this work took place in whole or in part at the Sanford Underground Research Facility (SURF) in Lead, South Dakota. Funding for this work is supported by the U.S. Department of Energy, Office of Science, Office of High Energy Physics under Contract Numbers DE-AC02-05CH11231, DE-SC0020216, DE-SC0012704, DE-SC0010010, DE-AC02-07CH11359, DE-SC0012161, DE-SC0014223, DE-FG02-13ER42020, DE-SC0009999, DE-NA0003180, DE-SC0011702, DESC0010072, DE-SC0015708, DE-SC0006605, DE-FG02-10ER46709, UW PRJ82AJ, DE-SC0013542, DE-AC02-76SF00515, DE-SC0019066, DE-AC52-07NA27344, \& DOE-SC0012447. This research was also supported by U.S. National Science Foundation (NSF); the U.K. Science \& Technology Facilities Council under award numbers, ST/M003655/1, ST/M003981/1, ST/M003744/1, ST/M003639/1, ST/M003604/1, and ST/M003469/1; Portuguese Foundation for Science and Technology (FCT) under award numbers PTDC/FIS-­PAR/28567/2017; the Institute for Basic Science, Korea (budget numbers IBS-R016-D1); University College London and Lawrence Berkeley National Laboratory thank the U.K. Royal Society for travel funds under the International Exchange Scheme (IE141517). We acknowledge additional support from the Boulby Underground Laboratory in the U.K., the GridPP Collaboration~\cite{faulkner:2006:gridpp,britton:2009:gridpp}, in particular at Imperial College London and additional support by the University College London (UCL) Cosmoparticle Initiative. This research used resources of the National Energy Research Scientific Computing Center, a DOE Office of Science User Facility supported by the Office of Science of the U.S. Department of Energy under Contract No. DE-AC02-05CH11231. The University of Edinburgh is a charitable body, registered in Scotland, with the registration number SC005336. The assistance of SURF and its personnel in providing physical access and general logistical and technical support is acknowledged.

\bibliography{LZ,mybibfile}

\end{document}